# Achromatic light patterning and improved image reconstruction for parallelized RESOLFT nanoscopy

Andriy Chmyrov[1,2,§†], Marcel Leutenegger[1†], Tim Grotjohann[1], Andreas Schönle[2], Jan Keller-Findeisen[1], Lars Kastrup[2], Stefan Jakobs[1,3], Gerald Donnert[2], Steffen J. Sahl[1] & Stefan W. Hell[1*]

[1] Max Planck Institute for Biophysical Chemistry, Department of NanoBiophotonics, Am Faßberg 11, 37077 Göttingen, Germany.
[2] Abberior Instruments GmbH, Hans-Adolf-Krebs-Weg 1, 37077 Göttingen, Germany.
[3] University of Göttingen, Medical Faculty, Department of Neurology, Robert-Koch-Str. 40, 37075 Göttingen, Germany.
[§] Present address: Helmholtz-Zentrum München, Institute of Biological and Medical Imaging, Ingolstädter Landstraße 1, 85764 Neuherberg, Germany.

[†] These authors contributed equally.

[*] Correspondence should be addressed to S.W.H. (hell@nanoscopy.de).

Fluorescence microscopy is rapidly turning into nanoscopy. Among the various nanoscopy methods, the STED/RESOLFT super-resolution family has recently been expanded to image even large fields of view within a few seconds. This advance relies on using light patterns featuring substantial arrays of intensity minima for discerning features by switching their fluorophores between 'on' and 'off' states of fluorescence. Here we show that splitting the light with a grating and recombining it in the focal plane of the objective lens renders arrays of minima with wavelength-independent periodicity. This colour-independent creation of periodic patterns facilitates coaligned on- and off-switching and readout with combinations chosen from a range of wavelengths. Applying up to three such periodic patterns on the switchable fluorescent proteins Dreiklang and rsCherryRev1.4, we demonstrate highly parallelized, multicolour RESOLFT nanoscopy in living cells at ~60–80 nm resolution for ~100×100 µm² fields of view. We discuss the impact of novel image reconstruction algorithms featuring background elimination by spatial bandpass filtering, as well as strategies that incorporate complete image formation models.

## Introduction

Scanning super-resolution[1,2] approaches including STED (STimulated Emission Depletion)[3] and RESOLFT (REversible Saturable/Switchable Optical Linear Fluorescence Transitions)[4,5,6] enjoy increasing popularity, in large measure due to their robust, direct image acquisition procedure. The STED/RESOLFT approach enables nanoscale recordings without the need for indirect, reciprocal-space data processing to obtain an image. In the STED/RESOLFT concepts, features located within subdiffraction length scales are directly discerned by transiently preparing their molecules in two distinct states (fluorescence 'on' and 'off') using a pattern of light featuring one or more intensity minima. STED nanoscopy has been applied for more than a decade, including to living organisms[7,8], providing resolutions down to 20 nm in biological specimens[9].

However, STED microscopy requires relatively large illumination intensities (MW/cm$^2$) because the transition from the fluorescent (on) to the dark (off) ground state by stimulated emission has to occur within the few nanoseconds lifetime of the on state. A successful strategy to reduce the required intensity is to select transitions between on- and off-states of longer lifetimes. Such states are provided by reversibly switchable fluorescent proteins (RSFPs), where the two states are caused by atom relocations within the fluorescent molecule (cis-trans-isomerization) or by addition of a water molecule. Hence, RSFPs can be switched with lower light intensities, commonly in the W/cm² to kW/cm² range. Although RESOLFT was initially introduced[6] as a general concept, it later became associated and was realized with RSFPs[10-12].

The initial demonstration of RSFP-based RESOLFT dates from 2005 (ref. 10), but only more recently has the development of new RSFPs allowed to provide proofs of concept of its promise as a nanoscale imaging tool in living cells at reduced light levels[11-16]. The low-intensity switching can be associated with reduced imaging speed. This limitation can be effectively counteracted by scanning with a pattern of standing waves that provides multiple intensity minima (parallelized RESOLFT)[17]. More recently, this scheme has also been extended to STED[18,19], but the need for high STED intensity limits the number of minima to a few thousands at a time. In parallelized RESOLFT, on the other hand, switching at low light intensities has rendered the implementation with ~100,000 simultaneous intensity minima possible. Like for the single-point scanning implementations of the methods, the demonstrated resolutions of parallelized RESOLFT[17] to date lag behind those of parallelized STED (20–30 nm)[19]. It is worth noting that the RSFPs used for parallelized RESOLFT were of the so-called negative-switching class. These are proteins for which the excitation light concurrently induces the on→off transition. As a result, the number of emitted fluorescence photons per switching cycle is limited. Further improvements to (parallelized) RESOLFT could therefore come from exploring RSFPs which do not suffer from this limitation. We here give an example of this, and describe highly relevant advances in both the light patterning required for switching and in the RESOLFT image reconstruction.

Specifically, we chose to develop parallelized RESOLFT capabilities for the previously introduced RSFP Dreiklang[12,21]. Distinct from all other RSFPs[20], all three processes, the on- and off-switching as well as fluorescence excitation of Dreiklang, are decoupled and individually addressable with light of different wavelengths. While this feature allows to tune the fluorescence signal per switching cycle and hence the signal-to-noise ratio (SNR) of the raw image data, its optimal implementation demands three co-aligned periodic light patterns of different wavelengths. These light patterns can be generated with shared phase gratings and achromatic projection optics such that their periods closely match in the sample. In a common beam path, we implemented patterned switching and fluorescence excitation of Dreiklang and the negative-switching RSFP rsCherryRev1.4[22,23] with laser light. However, since our projection optics had low transmittance of light at ~360 nm wavelength, we used a homogeneous LED illumination for switching Dreiklang on.

Parallelized RESOLFT implementations to date suffer from the blurry background stemming from out-of-focus features, as is well-known in wide-field microscopy. We addressed this issue by novel RESOLFT image reconstruction methods based on either a spatial bandpass filtering or a maximum likelihood fitting of the acquired images. This greatly improved the contrast of the RESOLFT images without sacrificing spatial resolution.

## Results

Parallelized RESOLFT nanoscopy using periodic standing-wave patterns can be realized in several ways, notably by an interferometer approach with reflective beam splitting[19] or by diffractive beam splitting using a diffraction grating[17,24], or in principle any other phase-changing device. In the interferometer approach a laser beam is first divided into two equal parts with a beam splitter. The individual beams are displaced from the optical axis of the microscope, recombined using a similar splitter and then focused on the back focal plane of the objective lens. It might look as if reflective splitting entails achromatic (wavelength-independent) beam paths due to identical reflection angles for different wavelengths, as well as direct beam control. However, chromatic dependence is unavoidable during beam recombination in the focal plane of the objective lens, where the beams are brought to interfere. The situation is different with diffractive beam splitting, where chromatic dependence is compensated for by the chromatic dependence of the interference at the sample.

We used two binary linear phase gratings with a bar-to-space ratio of 1:1 (0.5 duty cycle) and imaged them into the sample by transferring only their first diffraction orders. The phase gratings were oriented perpendicularly to each other and were lit by s-polarized light to maximize the contrast of the interference pattern in the sample.[1] Insofar as the imaging optics is achromatic, diffractive beam splitting and projection results in an interference pattern of equal period over a broad wavelength range. This is evident for microscopic imaging systems following the Abbe sine condition to achieve a uniform lateral magnification. For perpendicular incidence ($\theta_i = 0$), the first-order beams are diffracted by the phase grating under an angle $\sin(\theta_g) = \pm\lambda/\Lambda_g$, where $\lambda$ is the wavelength and $\Lambda_g$ the grating period. An imaging system following the Abbe sine condition transfers the beams into the sample at an angle $\sin(\theta_s) = \sin(\theta_g)/M_t$, where they form an interference pattern with an intensity period of $\Lambda_s = \Lambda_g|M_t|/2$ (**Fig. 1a**). We used phase gratings with $\Lambda_g = 36$ μm and set up a lateral demagnification $M_t = -1/50\times$ of the imaging optics. Thus, we obtained an illumination period of $\Lambda_s = 360$ nm.

**Supplementary Fig. 1** outlines the optical setup and its key components. The illumination patterns were imaged into the sample by two relay systems and the microscope optics. A first Keplerian telescope (lenses $L_2$ and $L_3$) was used to select the first diffraction orders. The order selection mask (OS) was inserted to block the non-diffracted zero-order light. A second Keplerian telescope (lenses $L_4$ and $L_5$) allowed to insert a tip-tilt scan mirror for shifting the illumination patterns without a need of moving the sample. **Figure 1b–d** shows images of the illumination patterns at 405, 488 and 592 nm wavelength, confirming the achromatic imaging of the gratings into the sample. The illumination periods at 488 and 592 nm wavelength differed by about 1 nm from the period at 405 nm wavelength, allowing for parallelized RESOLFT imaging in fields spanning up to ~300 periods (~100 μm) when accepting a shift of up to ~40% of a period between the illumination patterns. **Supplementary Fig. 13** shows the lateral shifts of the coordinates of the fluorescence peaks in the full image field of 104 μm. The illumination patterns for 405 and 488 nm wavelengths were imaged by the induced fluorescence of an Alexa Fluor 488 layer on camera 1. Therefore, the positions of the fluorescence peaks accurately report on the illumination patterns in the sample. However, the 592 nm illumination pattern excited a layer of ATTO590 whose fluorescence was imaged on camera 2. The larger and asymmetric shifts of the 592 nm pattern are partially due to differences in the aberrations of the two imaging paths. When imaging rsCherryRev1.4 we did not observe a significant modulation in brightness across the field of view that would indicate an insufficient match of the 592 and 405 nm patterns in the sample. Thus, we could comfortably image in fields of ~70 μm extent and up to ~100 μm when accepting a reduced image homogeneity in the border region.

Our parallelized RESOLFT setup can project illumination patterns in a wavelength range of ~400 to ~600 nm and can therefore operate with a large range of proteins requiring different wavelengths for switching and fluorescence excitation. In particular, it allows patterned fluorescence read-out, which in a previous realization of parallelized RESOLFT was performed using spatially uniform illumination[17]. Patterned read-out reduces the light exposure of the sample by illuminating mainly the sample regions where the molecules are left on after switching, i.e. the coordinates targeted by the minima of the off-switching pattern. Non-switched background and out-of-focus light is reduced as well while the effective lateral resolution is mostly determined by the off switching step (**Supplementary Fig. 14**).

---

[1] The interference of p-polarized light at high incidence angles would result in a strong z-polarized component that would partially fill the intensity minima.

Utilization of diffraction gratings offers further benefits such as increased opto-mechanical robustness and insensitivity versus wavelength drift. Higher opto-mechanical stability is achieved due to the common-path geometry used for projecting the diffracted beams into the sample. Also, any drift of the grating is demagnified 50×. In addition, laser beams with short coherence length can be used to create illumination patterns that are less perturbed by stray interference fringes from dust particles and optics irregularities.

We modified a commercial two-colour parallelized RESOLFT microscope (Abberior Instruments GmbH, Göttingen, Germany) and adapted it for Dreiklang[12]. The setup is described in detail in the supplementary methods. Dreiklang was switched on with spatially uniform UV-light from an LED source ($\lambda = 365$ nm, pE-4000, CoolLED, UK) coupled into the objective lens with a thin dichroic mirror (T387lp-UF1, AHF, Germany) transmitting wavelengths >390 nm. We recorded the images by scanning pairs of orthogonally overlapped standing-wave patterns for off-switching and for fluorescence read-out across the stationary sample. The fluorescence images collected by the objective lens were descanned so that the minima of the off-switching pattern remained stationary on the camera sensor. The camera recorded an image for each scan position. The pattern period was 360 nm for all wavelengths and was scanned in 10 or 14 steps along the $(x, y)$ directions, resulting in a pixel size of 36 or 26 nm in the reconstructed RESOLFT images.

Dreiklang was imaged by applying the following illumination sequence at each scan position: (i) spatially uniform illumination ($\lambda = 365$ nm; 0.2 kW/cm$^2$ for 15 ms) switched Dreiklang on; (ii) patterned near-UV light (MLD 405-200, Cobolt AB, Sweden; $\lambda = 405$ nm; 1 kW/cm$^2$ for 45 ms) switched it off everywhere except in the minima; and (iii) the proteins remaining on were probed by exciting fluorescence with patterned blue light (MLD 488-200, Cobolt AB, Sweden; $\lambda = 488$ nm; 1.5 kW/cm$^2$ for 12 ms) shifted by half a period relative to the off-switching pattern. The full 104 μm × 104 μm field of view was sampled in $14^2 = 196$ images within 17 seconds. Faster imaging can be achieved by shorter illumination periods at higher intensities. We deliberately chose a longer readout period than applied in point-scanning RESOLFT[19] to achieve better SNR in the raw images.

The acquired raw images showed significant background in regions with many sample features outside the imaged focal section. The previously published analysis method[17] estimated the in-focus and out-of-focus signals by integrating the image intensities in small pinholes placed at the positions of the minima and maxima of the off-switching pattern. These positions were identified as the periodic maxima and minima of the average image intensity. The local average out-of-focus signal was then subtracted from the in-focus signal to reconstruct the focal section at high resolution. However, due to the spatial extent of the in-focus detection point spread function (PSF), the correct identification of the in-focus and out-of-focus signal contributions is challenging.

We improved the image reconstruction over the previously published method[17] to more accurately identify the positions of the minima (nulls) of the off-switching illumination pattern and to better reject the fluorescence emission from out-of-focus features. These new analysis algorithms are described in detail in the **Supplementary Methods**.

The positions of the nulls were determined by analysing the average image intensity for the global $x$ and $y$ periods of the intensity peaks. Small misalignments and distortions of the illumination pattern were taken into account. We determined the local shifts of the nulls by correlating the periodic pattern with the fluorescence intensity peaks in neighbourhoods of 20 periods extent (**Supplementary Fig. 2**).

The rejection of the out-of-focus and non-switching background signal was addressed by one of two approaches: (a) A spatial bandpass filter was applied to the raw images to suppress non-switching and low-frequency signals
before integrating the signals in pinholes placed at the intensity peaks. We used a Gaussian smoothing of the raw images with a full width at half-maximum (FWHM) of $1/\sqrt{8}$ of a period to filter noise and then took the 2D Laplacian to enhance the signal variations. This combined linear filter acts on an image region of three periods extent and suppressed out-of-focus signals very efficiently. However, as the filter response is negative in an annulus around the central peak, it may suppress also the in-focus signals in the immediate vicinity of a very bright feature. We removed this potentially artifactual effect to a large degree by reattributing the signals according to the response of the bandpass filter and the pinholes (**Supplementary Fig. 3**).

(b) Based on the known positions of the potentially highly emissive regions in the sample, we applied an image formation model for estimating the various signal contributions to the raw images. The image formation model accounted for contributions from several optical sections – the focal section and two out-of-focus sections located further from the cover slip – by estimating for each null and each section the fluorescence signal as a model coefficient. The model itself consisted of the detection PSFs for fluorescence emission at these positions in the sample (**Supplementary Fig. 4**). The PSFs were calculated in the pixilated sensor matrix individually for each null applying calculation tools developed by Leutenegger et al.[25-27]. We used an iterative conjugate gradient fitting algorithm to maximize the likelihood of observing the acquired raw images given the model and its coefficients.

Examination of recorded RESOLFT images of Dreiklang fused to keratin19 (**Fig. 2**) shows that thin filaments can be identified with improved resolution while they appeared blurred in the corresponding wide-field images. **Figure 2a** and **Supplementary Fig. 6** show a reconstructed wide-field image as summed from the acquired images by resampling each image at the RESOLFT pixel size and at its true position. **Figure 2b** and **Supplementary Fig. 7** illustrate the contrast enhancement obtained by deconvolving the wide-field image with the theoretical detection PSF. Both images show poor contrast due to high background from many defocused sample features. The FWHM of individual filaments is about 220 nm, and so is the diffraction-limited spatial resolution.

**Figure 2c** and **Supplementary Fig. 8** illustrate the RESOLFT image quality as obtained with the previously published analysis, but already amended by the local estimation of the nulls' positions. The background was estimated as 80% of the average signal in between the diagonal neighbours of each null. The residual stripes are attributed to a different photo-bleaching behaviour of the (non-switching) background versus the signals at the nulls. This method could not very reliably disentangle the background from the true signal, as evidenced by the still pronounced presence in all regions, where the wide-field image shows notable out-of-focus emission.

**Figure 2d** and **Supplementary Fig. 9** illustrate the RESOLFT image quality with the spatial bandpass filtering pre-processing. The image contrast is improved significantly throughout. Note that some dark regions are produced around the very bright features in the cell on the top right of the image (displayed as saturated pixels). In these regions, the bandpass filtered images show negative values which were not fully compensated by the reattribution step.

The imaging model-based RESOLFT reconstruction efficiently distinguished between the emissions from out-of-focus features including potential non-switchable background and the wanted emission from in-focus features. **Figure 2e** and **Supplementary Fig. 10** show the in-focus image obtained by analysing the experiment with the imaging model. The focused features are visible with high contrast on a negligible background. The quality of the RESOLFT image of this measurement remained limited by the moderate SNR, and neither the bandpass filtering nor the model-based analysis improved the lateral resolution as compared to the previous analysis method. To compare the contrast and resolutions of the five imaging methods, **Fig. 2f–i** zoom into an interesting region. A line profile crossing three or four filaments shows that the contrast improved by more than an order of magnitude from the wide-field image to the model-based RESOLFT reconstruction. Individual filaments were typically imaged with 60–80 nm FWHM as determined by fitting a Lorentzian function to cross-sections (**Supplementary Fig. 11**, 36 profiles averaged over 5 pixels (130 nm) along the filaments). The bundle of (likely three) filaments and one separate filament crossing the line profile were resolved at ~60 nm and 120 nm mutual distances (**Fig. 2j**), where the wide-field and deconvolved images show a single feature of ~300 nm FWHM. The analysis of closely spaced, nearly parallel filaments was mostly hampered by other filaments crisscrossing above or beneath. It is worthwhile to note for interpretation that keratin protein bundles to form intermediate filaments that are of varying diameters, and that it is consequently the thinnest filaments which are most suitable for resolution estimates.

We further demonstrate parallelized RESOLFT nanoscopy with Dreiklang fused to the microtubule associated protein MAP2 (**Fig. 2k**,**l** and **Supplementary Fig. 12a**,**b**) and the endoplasmic reticulum (**Fig. 2m**,**n** and **Supplementary Fig. 12c**,**d**).

The achromatic grating projection simplified the acquisition of multi-colour images using green and red RSFPs, such as Dreiklang and rsCherryRev1.4[23]. For these proteins, we found it challenging to image both RSFPs simultaneously or in an interleaved manner, because the relatively long irradiation with 405 nm light to switch Dreiklang off weakened the rsCherryRev1.4 signal by photobleaching. Therefore, we imaged Dreiklang after the rsCherryRev1.4 images had been taken. Each spectral range was acquired by a dedicated sCMOS camera (Hamamatsu ORCA-Flash4.0 V2). The dual-colour images are shown in **Fig. 3.** For the dual-colour images, rsCherryRev1.4 was imaged in 10×10 scan steps: (i) on-switching with 405 nm light at 0.2 kW/cm$^2$ for 1.5 ms; (ii) off-switching with 592 nm light at 0.9 kW/cm$^2$ for 50 ms; and (iii) read-out with 592 nm light at 0.9 kW/cm$^2$ for 10 ms. Dreiklang was imaged using the same scan steps: (i) uniform on-switching with 365 nm light at 0.2 kW/cm$^2$ for 15 ms; (ii) off-switching with 405 nm light at 0.2 kW/cm$^2$ for 80 ms; and (iii) read-out with 488 nm light 0.8 kW/cm$^2$ for 10 ms.

## Discussion

We demonstrated that the diffractive pattern generation and imaging adopted in our parallelized RESOLFT nanoscope is sufficiently achromatic to allow for illuminations at several wavelengths with light patterns of equal periodicity. Essentially any fluorescent marker – with arbitrary requirements on excitation, switching and emission spectra – could be utilized by feeding the respective laser wavelengths to the illumination system. This will readily facilitate the adoption of new fluorescent proteins as they become available (e.g. ref. 28-30). We showed that the resolution of parallelized RESOLFT can be increased to 60–80 nm in living cells expressing the RSFP Dreiklang. In terms of applicability to other RSFPs with specific wavelength requirements, parallelized RESOLFT might be limited by residual chromatic aberrations. On the hardware side, inexpensive continuous-wave (cw) lasers of sufficient power are available in the entire visible wavelength spectrum, readily facilitating the adoption of new switchable fluorophores.

We showed novel image reconstruction analyses that significantly improve RESOLFT image contrast by efficiently eliminating out-of-focus background. Spatial bandpass filtering of the raw images, at little computational cost, can analyse an acquired image sequence in about a minute. Being essentially an edge enhancer, a similar spatial bandpass filter cannot be applied to the wide-field image as it would break up continuous features. At the cost of significant computation time (~30s per raw image), our imaging-model-based analysis further decomposes the raw image contributions into fluorescence emissions from predefined regions with clearly different PSF sections – which distinguishes it from deconvolution algorithms that assume a location-independent PSF and/or do not restrict the locations of signal origins. Without knowledge of the restricted set of positions of fluorescence emissions, it is hardly possible to decompose a raw image into contributions from different sections, because a blurred image of an out-of-focus object could always be interpreted as an extended object in the focal plane. In contrast, knowing the positions and their sparse distribution in the sample allows to accurately attribute the origins of the different signals making up the raw images. In order to find the coefficients reliably, the raw images must provide redundant information. As the illumination patterns were imaged by the camera sensor with ~3.5 pixels per period, there were about twelve pixels per null. A decomposition into three coefficients per null seemed reasonable as it still offered some redundancy for dealing with noise.

Particular RSFP combinations may enable additional information channels per camera sensor, e.g. two colour channels sharing the same emission spectrum based on the opposite switching action of 405 nm light on Dreiklang and an off-switching RSFP such as rsEGFP[13,15]. Both proteins can be excited and read out at 488 nm wavelength, but 405 nm light switches Dreiklang off and rsEGFP on. The practical limitation of such an imaging scheme is some off-switching background from rsEGFP, which will result in cross-talk into the Dreiklang channel. Such photochromic channel separation could conceivably be combined with genuine spectral separation like the presented Dreiklang+rsCherryRev1.4 combination to implement tricolour parallelized RESOLFT imaging. As their properties are crucial for the image formation and RESOLFT image quality, improving the proteins in terms of brightness, switching contrast and speed will augment this versatile nanoscopy method further.

## Author contributions

The project originated from discussions between A.C. and S.W.H. A.C. designed and constructed the optical setup, performed imaging experiments. M.L. and A.S. contributed simulation and analysis tools, including the software Imspector (A.S.). A.C. and M.L. analysed data. T.G. developed biological samples for imaging demonstrations. L.K. advised and helped with construction of the optical setup. S.J. consulted on and provided fluorescent proteins. A.C., J.K.-F., M.L. and S.J.S. discussed the data, improvements to the analysis and the conclusions. A.C., M.L., S.J.S. and S.W.H. wrote the manuscript. All authors commented on the final version of the manuscript.

## Acknowledgements

We acknowledge funding from the German Federal Ministry of Education and Research (BMBF) within the project Ultraschnelle RESOLFT-Mikroskopie (VideoRES) (FKZ: 13N12993, 13N12995).

## Competing Financial Interests

A.S., L.K., G.D. and S.W.H. have personal financial interest (ownership) in Abberior Instruments GmbH, Germany. A.C. was employed by this company during this project, T.G. is now employed by this company. The concepts demonstrated in this publication find application in commercially available microscope systems.

## Methods

**Raw images acquisition.** Two-dimensional sets of coordinates were established by the periodic light patterns switching fluorescent proteins on and off. Fluorescence was excited by patterned illumination. The tip-tilt scan mirror projected the sequence of illumination patterns at the desired coordinates. The induced signals were imaged simultaneously on camera sensors. The illumination patterns were scanned in multiple steps to cover the entire unit cell of the periodic pattern such that all positions in the sample were addressed. The fluorescence was descanned to ensure that the signals within a unit cell were always imaged at the same detector pixels, facilitating the correction for pixel-specific non-uniformities (converter gain and noise, dark current) and the image reconstruction. The acquired raw images were further processed to attribute the signals to their respective coordinates (**Supplementary Figs. 2–4**, **Supplementary Methods**).

**Correcting the pattern displacement.** The lateral offsets between patterns of different wavelengths were determined separately along the horizontal and vertical directions. During the measurements for either component, the respective other was blocked (no light in orthogonal pattern component). To establish the offset between 405 nm and 488 nm patterns (**Supplementary Fig. 5**), pairs of images of dense keratin19–Dreiklang structures in HeLa cells were acquired. For each image, Dreiklang was switched on by 365 nm light, then switched off by patterned 405 nm light and finally imaged by patterned 488 nm light. The first image of each pair was acquired by displacing the read-out pattern by 180 nm with respect to the off-switching pattern. The second image of each pair was acquired by other pattern displacements in steps of 10 nm from 0 to 360 nm to cover a full period. The mean intensities of the second images were then normalized by the mean intensities of the first images to account for photo-bleaching. The normalized fluorescence was therefore modulated due to variable spatial coincidence of off-switching and read-out, being minimal when the two patterns were completely overlapping, and maximal when completely out-of-phase. Hence, the relative offset between the patterns of two wavelengths was determined and taken into account for subsequent measurements. The same strategy was used with rsCherryRev1.4 for measuring the lateral offset between 405 nm (on-switching) and 592 nm (off-switching and read-out) illumination patterns. Due to the low dose needed for switching on and the good switching contrast of Dreiklang and rsCherryRev1.4, the offset of the on-switching versus the off-switching pattern was less crucial here than the offset of the read-out versus the off-switching pattern.

**Fluorescent layers.** For direct imaging of the pattern with 405 nm and 488 nm light, a sample with a thin layer of Alexa Fluor 488 was created. The surface of a standard microscope cover glass (#1.5) was functionalized with the amine group via amino-silanization chemistry. After that, NHS-ester of Alexa Fluor 488 (Life Technologies, USA) was conjugated to the amine groups, producing a thin and close to uniform layer of the fluorophore. For the imaging of the pattern with 592 nm light, we diluted the fluorophore ATTO590 (ATTO-Tec, Germany) to 1 μM concentration in polyvinyl alcohol (PVA, Sigma-Aldrich, USA) and spun 150 μL of the solution onto standard (#1.5) microscope cover glasses (4000 rpm, 30s). The cover glasses with the fluorophore layer were mounted onto microscope glass slides using 20 μL DPX mounting medium (Sigma-Aldrich, USA).

**Expression of fusion proteins in HeLa cells.** To express protein fusions of Dreiklang with keratin19, MAP2 and the endosplasmatic reticulum's retention signal KDEL, previously reported plasmids were used[12]. HeLa cells were cultured at 37°C and 5% $CO_2$ in DMEM (Invitrogen, Carlsbad, USA) containing 10% FBS (PAA Laboratories, Cölbe Germany), 1mM pyruvate (Sigma, St.-Louis, USA) 100μg/ml streptomycin and 100μg/ml penicillin (Biochrom, Berlin, Germany). For transfection, cells were seeded on coverslips in a 6-well plate. After one day cells were transfected with Lipofectamine 2000 (Life Technologies, Carlsbad, USA) according to the manufacturer's instructions. 24 to 48 hours after transfection cells were washed and mounted with phenol-red free DMEM (Invitrogen, Carlsbad, USA) onto concavity slides. To prevent samples from drying, the coverslips were sealed with twinsil (Picodent, Wipperfürth, Germany). Finally the cells were imaged at room temperature. For two-colour experiments keratin19–rsCherryRev1.4 was additionally expressed using the plasmid pMD-Ker19-rsCherryRev1.4[23].

# Figures

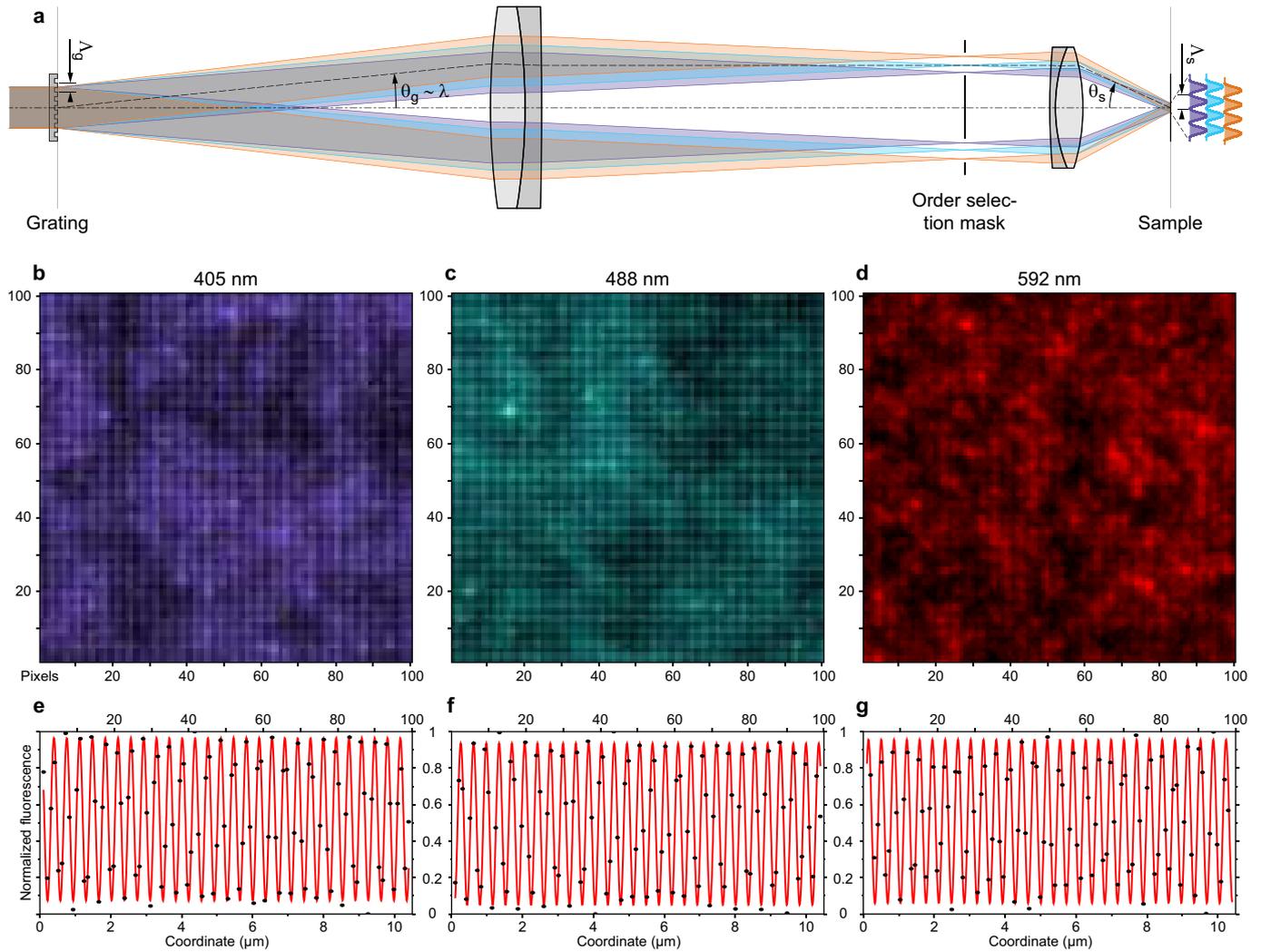

**Figure 1 | Achromatic behaviour of the parallelization scheme.** (**a**) Because the phase grating is imaged into the sample plane, the period of the interference patterns ($\Lambda_s$) is independent of the wavelength ($\lambda$) of light incident on the grating and equal to half the grating period ($\Lambda_g$) times the lateral magnification ($M_t$) of the imaging system, while the angle of diffraction from the grating ($\theta_g$) is proportional to the wavelength. (**b–d**) Fluorescence response from thin dye layers excited by light patterns at 405, 488 and 592 nm wavelength. (**e–g**) Intensity profiles of fluorescence for the three cases (the horizontal pattern component is shown, with the signal summed along the vertical direction and Fourier-filtered to remove low-frequency variation in fluorophore concentration, black dots), revealing equal periodicity but different offsets (red lines). The offset between the patterns of different wavelengths is determined by the phase difference due to chromatic variation. The offsets were measured and taken into account (see **Supplementary Fig. 5**). See **Supplementary Fig. 13** for an illustration of the pattern mismatches in the full field of view.

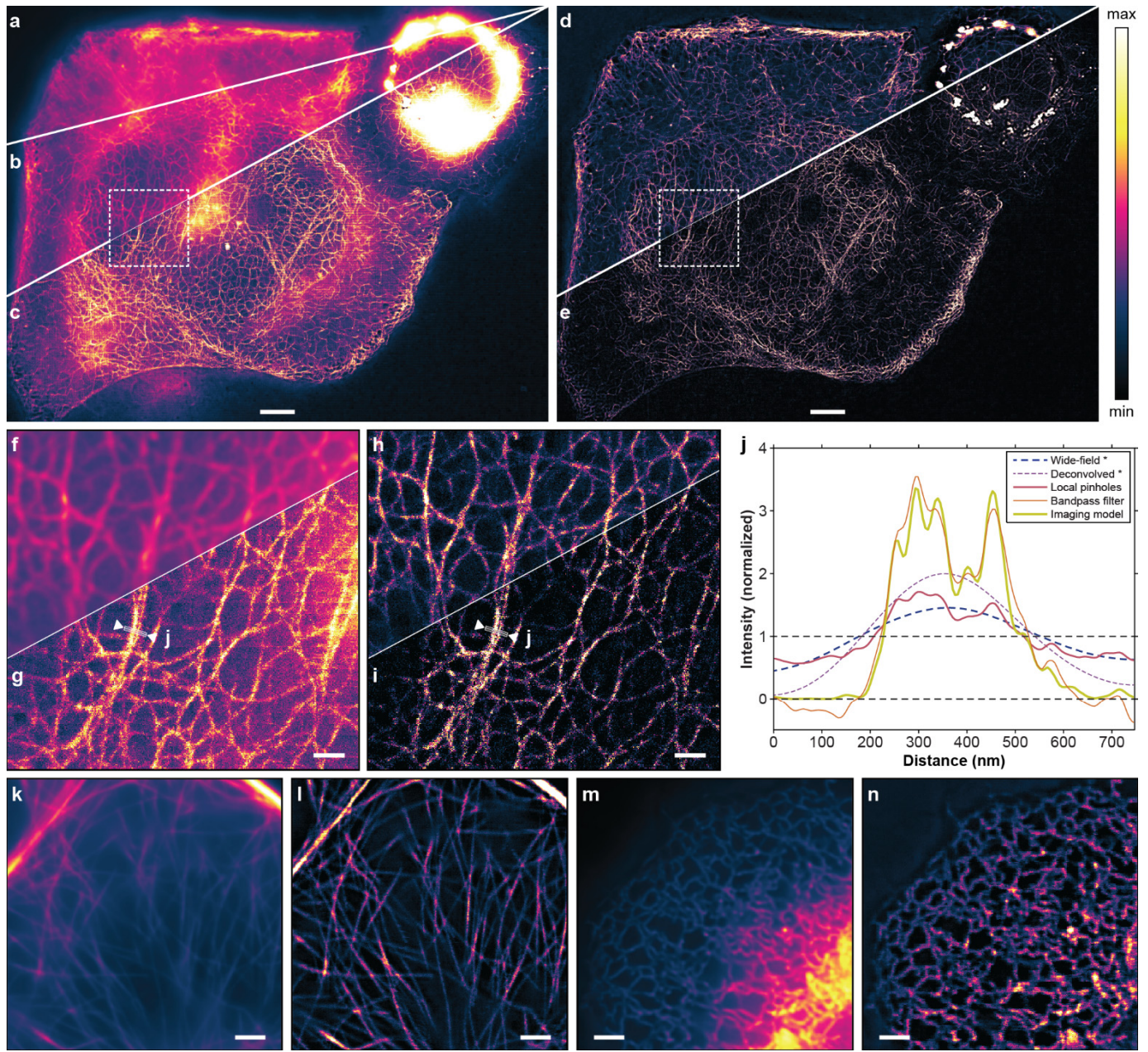

**Figure 2 | Parallelized RESOLFT nanoscopy of living HeLa cells expressing proteins fused to Dreiklang.** In about 17 seconds, 14×14 raw camera frames were acquired in 89 ms each and subsequently analysed with five methods. (**a**) Wide-field image composed from the 196 realigned raw camera frames (full image in **Supplementary Fig. 6**) and (**b**) deconvolved wide-field image (full image in **Supplementary Fig. 7**). RESOLFT images reconstructed with local pinholes (**c**), by bandpass filtering (**d**) and by applying the image formation model (**e**) (full images in **Supplementary Figs. 8–10**). The outlined regions in **b,c** and **d,e** are displayed in (**f,g**) and (**h,i**) magnified 4.5×, respectively. (**j**) Line profiles indicated in **g,i** and averaged over a width of 5 pixels (130 nm) for the five analysis methods. The profiles were normalized by their average values. *The contrast of the profiles of the images **a,b** around the average values is shown enlarged 4× for clarity. (**k–n**) Magnified region (wide-field images (**k,m**) and RESOLFT reconstruction with bandpass filtering (**l,n**)) of a HeLa cell expressing MAP2–Dreiklang (**k,l**) and KDEL–Dreiklang (**m,n**); full images in **Supplementary Fig. 12**). Scale bars: 5 µm (**a–e**), 1 µm (**f–i**), 2 µm (**k–n**). Displayed fields of view: 78×61 µm² (**a–e**), 11×11 µm² (**f–i**), 18×18 µm² (**k–n**).

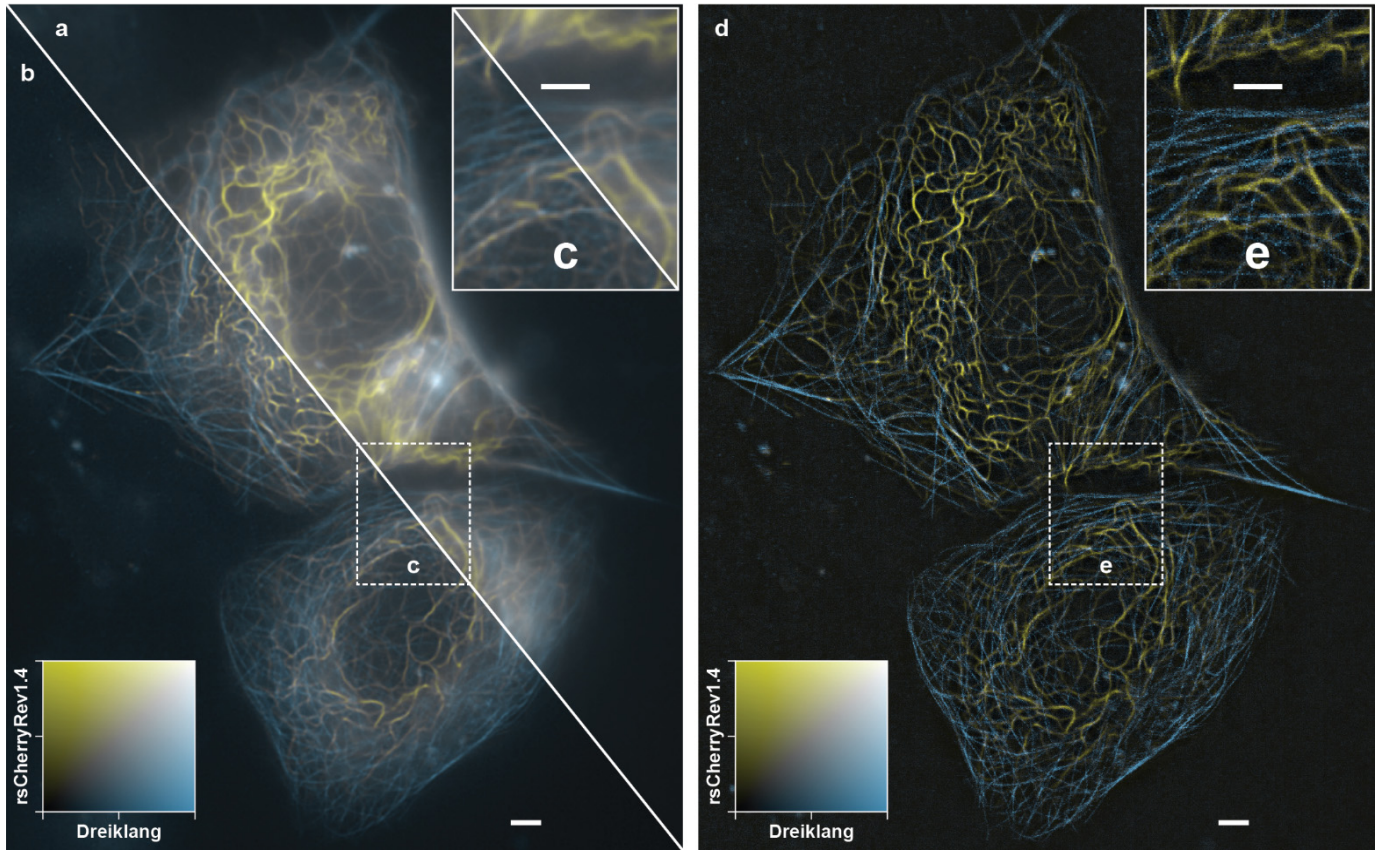

**Figure 3 | Two-colour live-cell imaging with parallelized RESOLFT nanoscopy.** Wide-field (**a**), deconvolved wide-field (**b**) and bandpass-filtered two-colour RESOLFT image (**d**) of live HeLa cells expressing keratin19–rsCherryRev1.4 and MAP2–Dreiklang. Displayed field of view: 56×70 µm². (**c**,**e**) 2× magnified regions. The RESOLFT images were taken with 10×10 scan steps. Each raw camera frame was acquired in 70 ms for keratin19–rsCherryRev1.4 and 113 ms for MAP2–Dreiklang, resulting in total imaging times of about 7 plus 11 seconds. All scale bars: 2 µm.

# Supplementary Figures

| | |
|---|---|
| $L_1$ | Achromat f = 40 mm (Qioptiq G322209000) |
| $L_2$ | Achromat f = 100 mm (Thorlabs AC254-100-A) |
| $L_{3,4}$ | Achromat f = 125 mm (Thorlabs AC254-125-A) |
| $L_5$ | Achromat f = 200 mm (Thorlabs AC254-200-A) |
| $L_{6,7,8}$ | Achromat f = 160 mm (Qioptiq G322352000) |
| | |
| $D_1$ | Dichroic 390/485/593 nm bandpass (Chroma ZT390/485/593tpc-UF5) |
| $D_2$ | Dichroic 580 nm edge BrightLine (Semrock FF580-FDi01-25x36) |
| $D_3$ | Dichroic 387 nm long pass (Chroma T387lp-UF1) |
| $F_1$ | Bandpass filter 550/88 nm BrightLine (Semrock FF01-550/88-25) |
| $F_2$ | Bandpass filter 641/75 nm BrightLine (Semrock FF02-641/75-25) |
| | |
| $G_{1,2}$ | Binary phase grating, 36 µm period, 1:1 duty cycle (Laser-Laboratorium Göttingen) |
| PBS | Polarizing beam splitter (B. Halle PTW 25) |
| λ/2 | Achromatic half-wavelength retarder plate (Thorlabs AHWP05M-600) |
| OS | Order selection mask, 4× Ø1 mm hole on Ø3 mm circle |

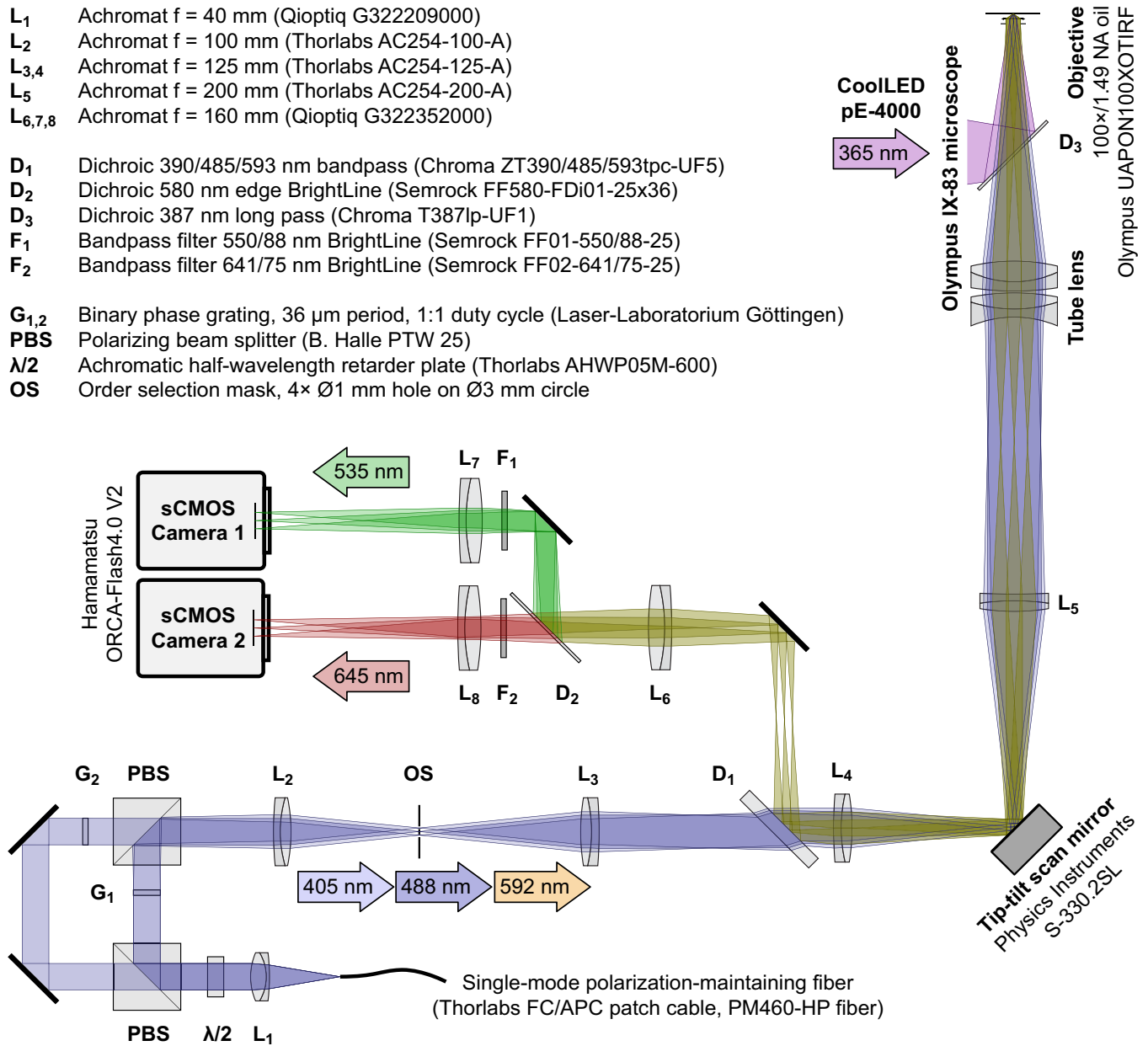

**Supplementary Figure 1 | Scheme of the two-colour parallelized RESOLFT setup.** The microscope features three patterned illuminations at 405, 488 and 592 nm wavelengths; a homogeneous illumination at 365 nm wavelength; and two fluorescence detection channels with 500–575 nm and 610–680 nm wavelength ranges.

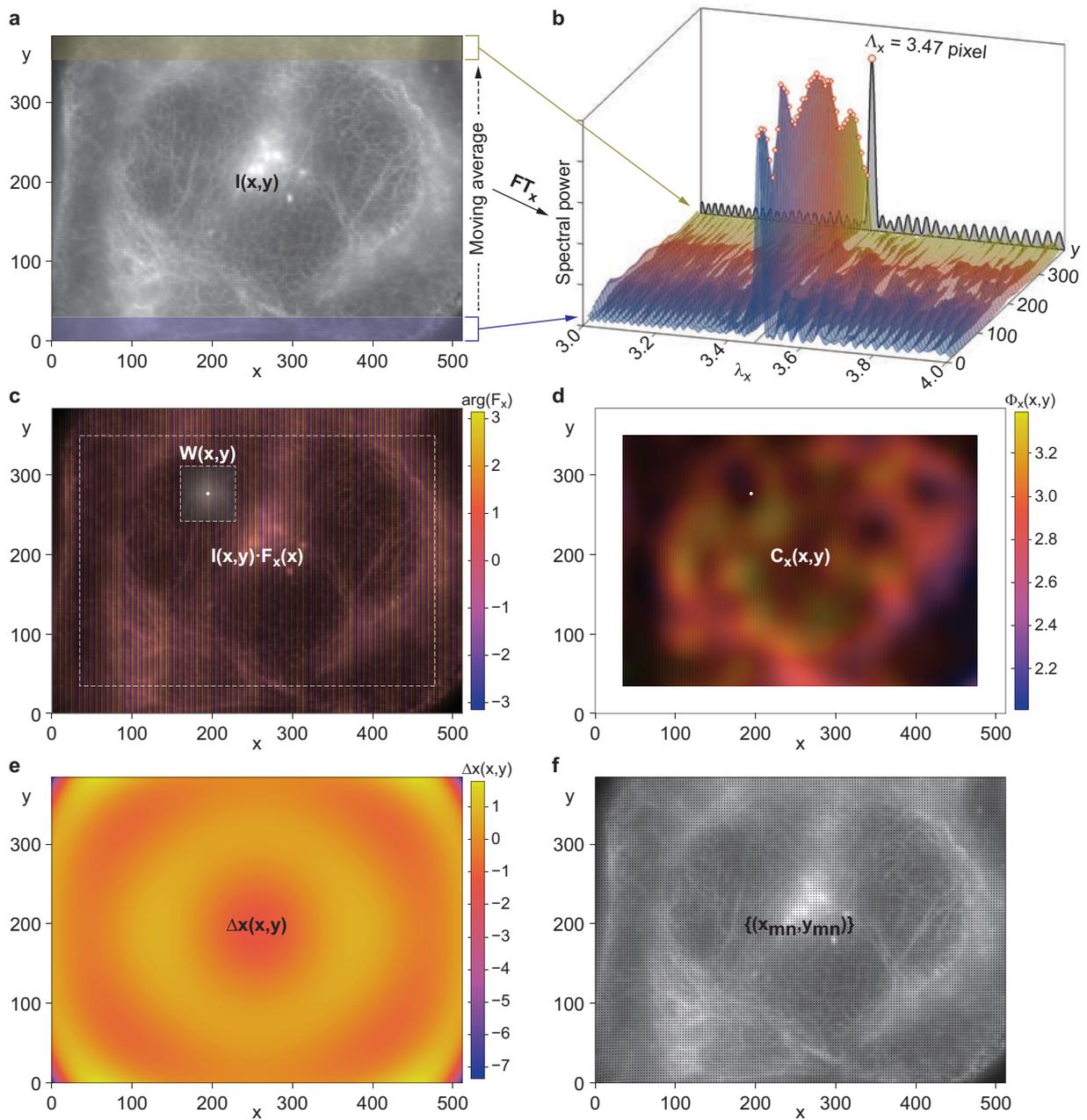

**Supplementary Figure 2 | Localization of the signal origins.** The illumination periods are estimated globally and the position offsets locally. (**a**) Sum image $I(x,y)$ and regions of the profiles $I_x(x,y)$ whose spectral powers (**b**) are analysed to find the horizontal period $\Lambda_x$. (**c**) Image $I(x,y)$ modulated by $F_x(x)$ and exemplary correlation window $W(x,y)$ with central point. The modulation phase is encoded as colour and the image intensity as brightness. (**d**) Local correlations $C_x(x,y)$. The unwrapped correlation phase $\Phi_x(x,y)$ is encoded as colour and the correlation amplitude as brightness. (**e**) Zernike-filtered and -extrapolated local position offset $\Delta x(x,y)$. (**f**) Image overlaid with the localized signal origins $X = \{(x_{mn}, y_{mn})\}$ (black dots). Units: $x, y, \lambda_x, \Lambda_x$ and $\Delta x$ in pixels; phases $\arg F_x$ and $\Phi_x$ in radians.

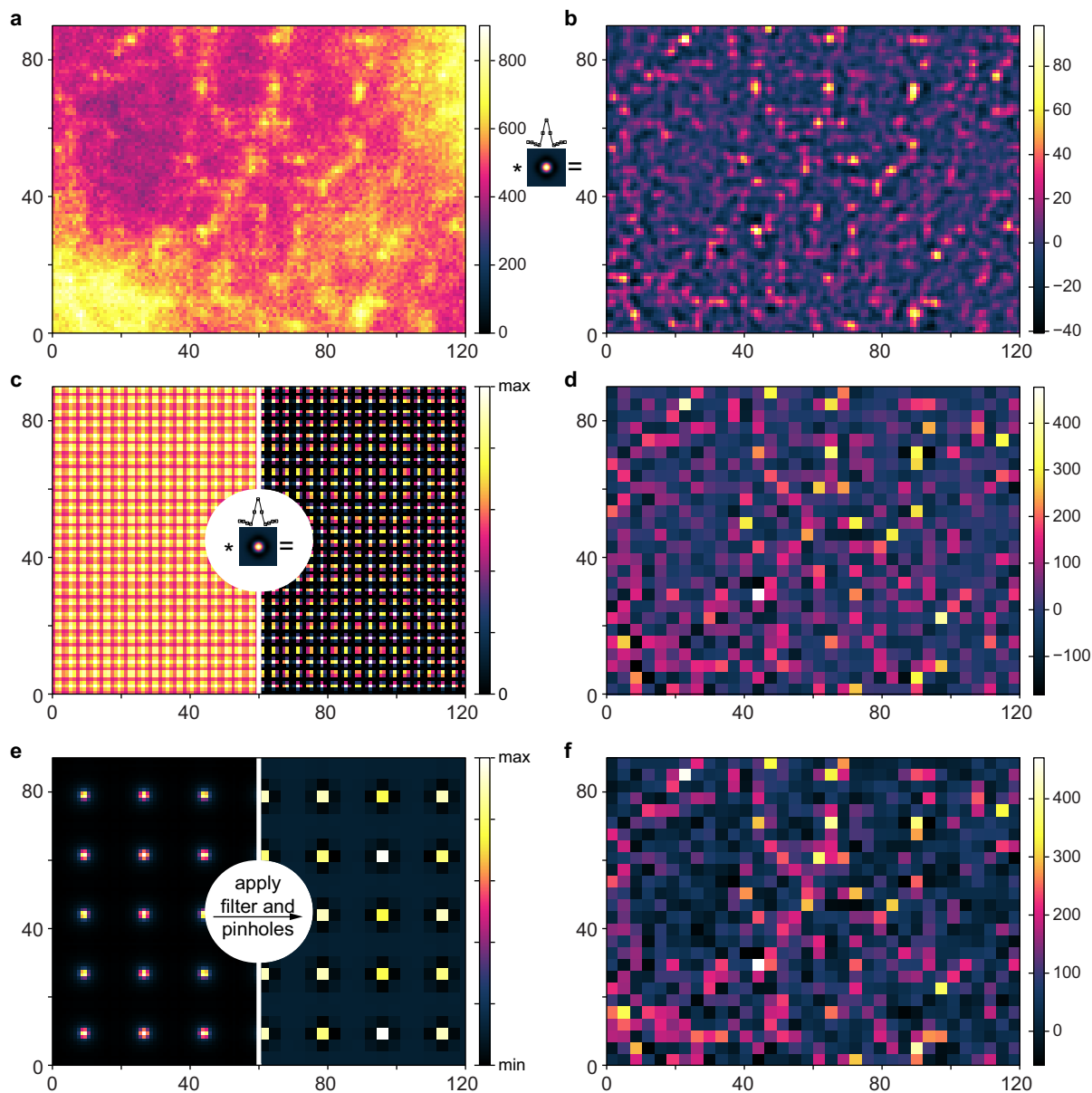

**Supplementary Figure 3 | Spatial bandpass filtering and the extraction of the signals emitted at the positions of the nulls.** (**a**) An acquired camera image in a small region of 120×90 pixels (about 35×26 illumination periods). The analog-to-digital converter (ADC) bias was subtracted and the ADC conversion factor was applied to scale the image in photo-electrons. (**b**) Bandpass filtered image obtained by convolving the acquired image **a** with the Laplacian-of-Gaussian filter kernel shown to scale in-between the images. (**c**) Determination of the pinholes and sensitivities. Left: Simulated image of equally bright emitters at every position of a null. Right: Bandpass filtered simulated image with applied zero threshold. Positive values are identifying the pinhole regions (4–6 pixels within a 3×3 neighbourhood at the nulls). (**d**) Extracted signals at the nulls obtained by the integrated signal in each pinhole normalized with its sensitivity. (**e**) The pinhole PSFs are extracted from simulated and analysed images. Equally bright emitters are placed at every fifth null to build all PSFs with only 5×5 images. (**f**) The PSF matrix is then applied in a least squares conjugate gradient fit to find the closest matching pinhole signals (shown) that yield the observed image **d**.

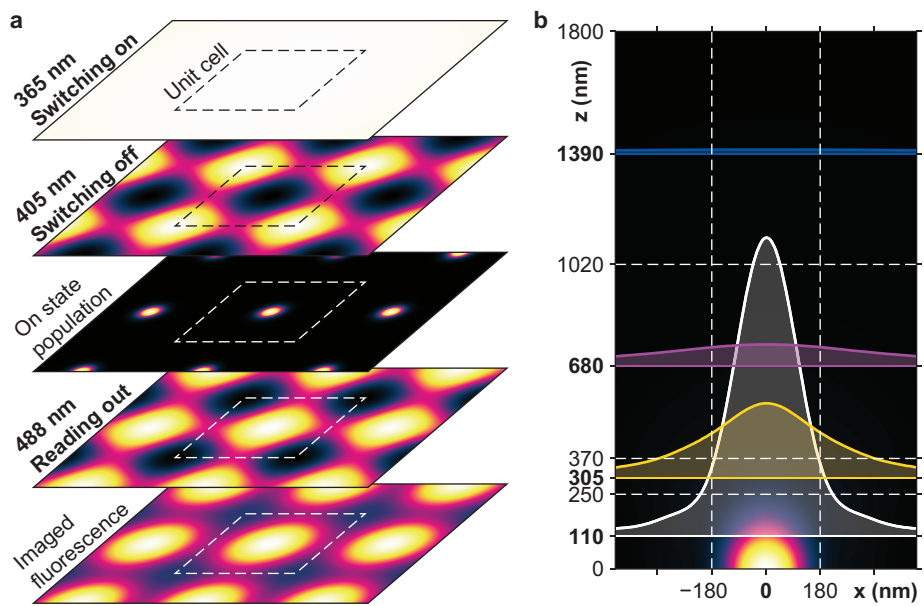

**Supplementary Figure 4 | Imaging PSF for the model-based RESOLFT image reconstruction.** (**a**) Estimation of the fluorescence image based on the illumination patterns and illumination doses when switching and exciting Dreiklang. The on state population and the read-out energy determine the potential fluorescence emission. The lowest panel illustrates the fluorescence image in the focal plane. (**b**) Estimation of the imaging PSF and separation into a small number of axial sections. The focal section extended from the cover-slip interface ($z = 0$ nm) into the sample such that its PSF area was $\sqrt{2}\times$ the focal PSF area at cover-slip interface. Adjacent sections were placed such that their PSF areas quadrupled from section to section. Dashed lines indicate the section borders and the lateral extent of the unit cell. The mutually distinct section PSFs are illustrated by cross-sections.

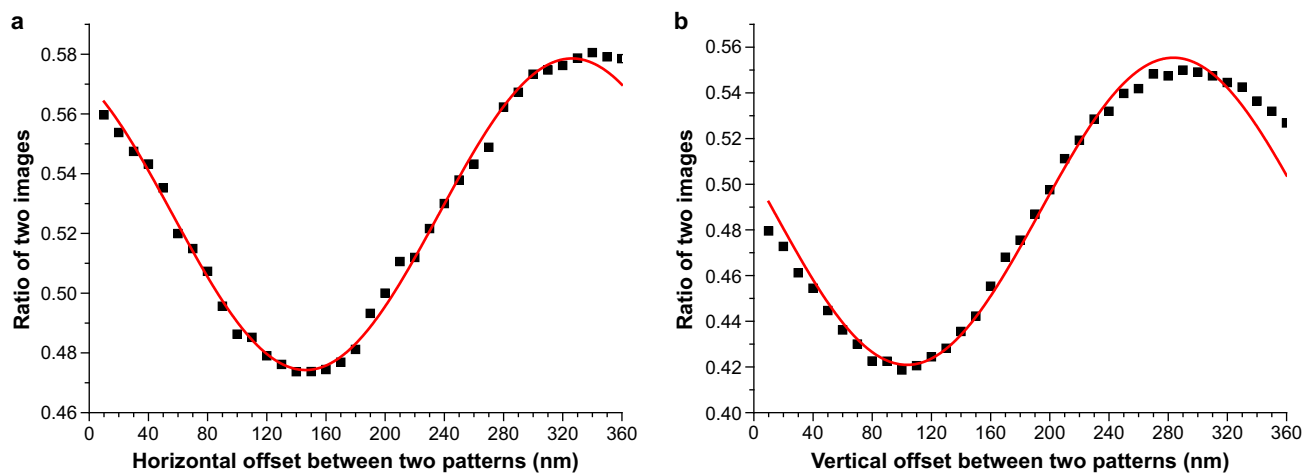

**Supplementary Figure 5 | Determination of offset between patterns due to phase shift.** The offsets between patterns of 405 and 488 nm light for horizontal (**a**) and vertical (**b**) components are determined separately by blocking the other component (no light in the orthogonal pattern). A sequence of two images each, of dense keratin19–Dreiklang structures, were taken, with activation, off-switching and read-out illuminations. For the first image, the pattern was displaced by half a period (180 nm) between the off-switching and the read-out step, as would be suitable if these patterns overlapped completely. For the second image, an additional variable shift of 10–360 nm in steps of 10 nm was introduced. Fluorescence in the recorded images was therefore modulated due to variable spatial coincidence of switch-off and read-out, with it being minimal when the two patterns (off-switching and read-out) were completely overlapping, and being maximal when completely out-of-phase. From this ratio we determined the relative offset between the patterns of two wavelengths and took it into account for later measurements. Black dots represent the ratios of the mean fluorescence in the images with the offset to the mean fluorescence in the images without the offset. Red lines represent the fit to sinusoidal functions with 360 nm period. The maximal ratio was found at the optimal offsets of 320 nm (**a**) and 280 nm (**b**).

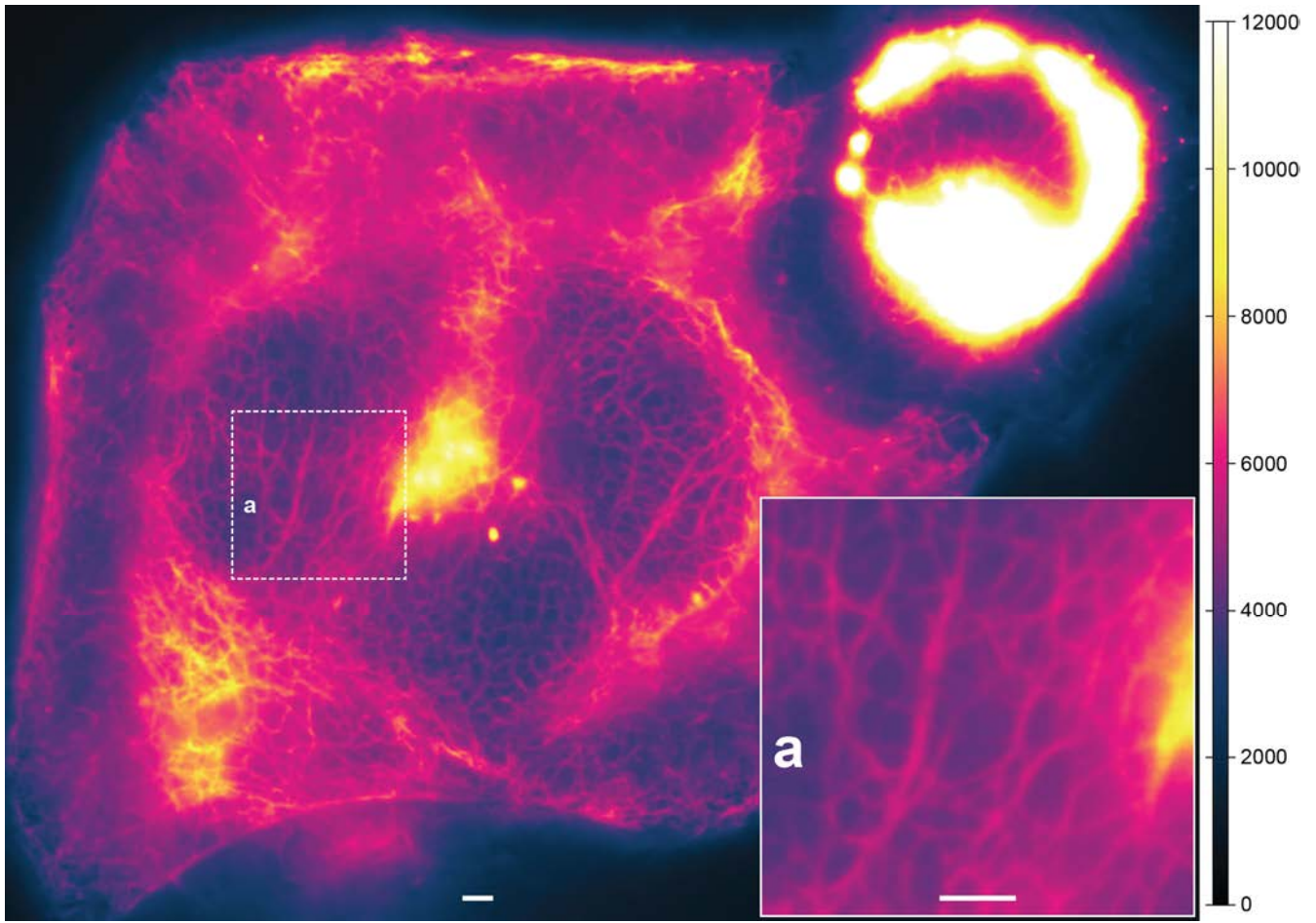

**Supplementary Figure 6 | Wide-field image obtained by shifting and summing the acquired images.** The image brightness was saturated at a maximum of 12'000 for clarity (peak brightness 121'000). (**a**) 2.5× magnified region. Scale bars: 2 μm. Displayed fields of view: 78×61 μm$^2$, 11×11 μm$^2$ (**a**).

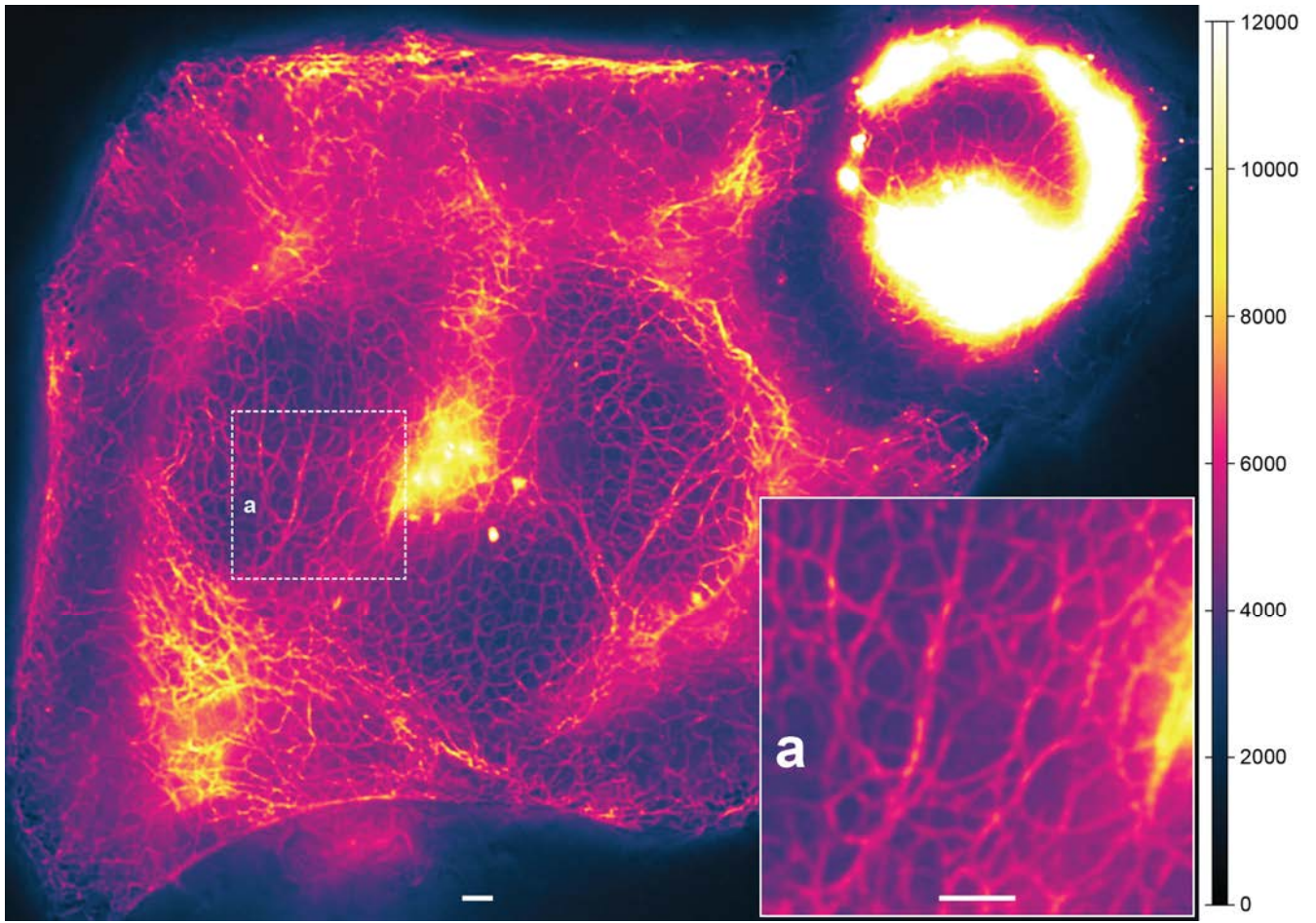

**Supplementary Figure 7 | Deconvolved wide-field image as shown in Supplementary Figure 6.** The image brightness was saturated at a maximum of 12'000 for clarity (peak brightness 153'000). (**a**) 2.5× magnified region. Scale bars: 2 µm. Displayed fields of view: 78×61 µm$^2$, 11×11 µm$^2$ (**a**).

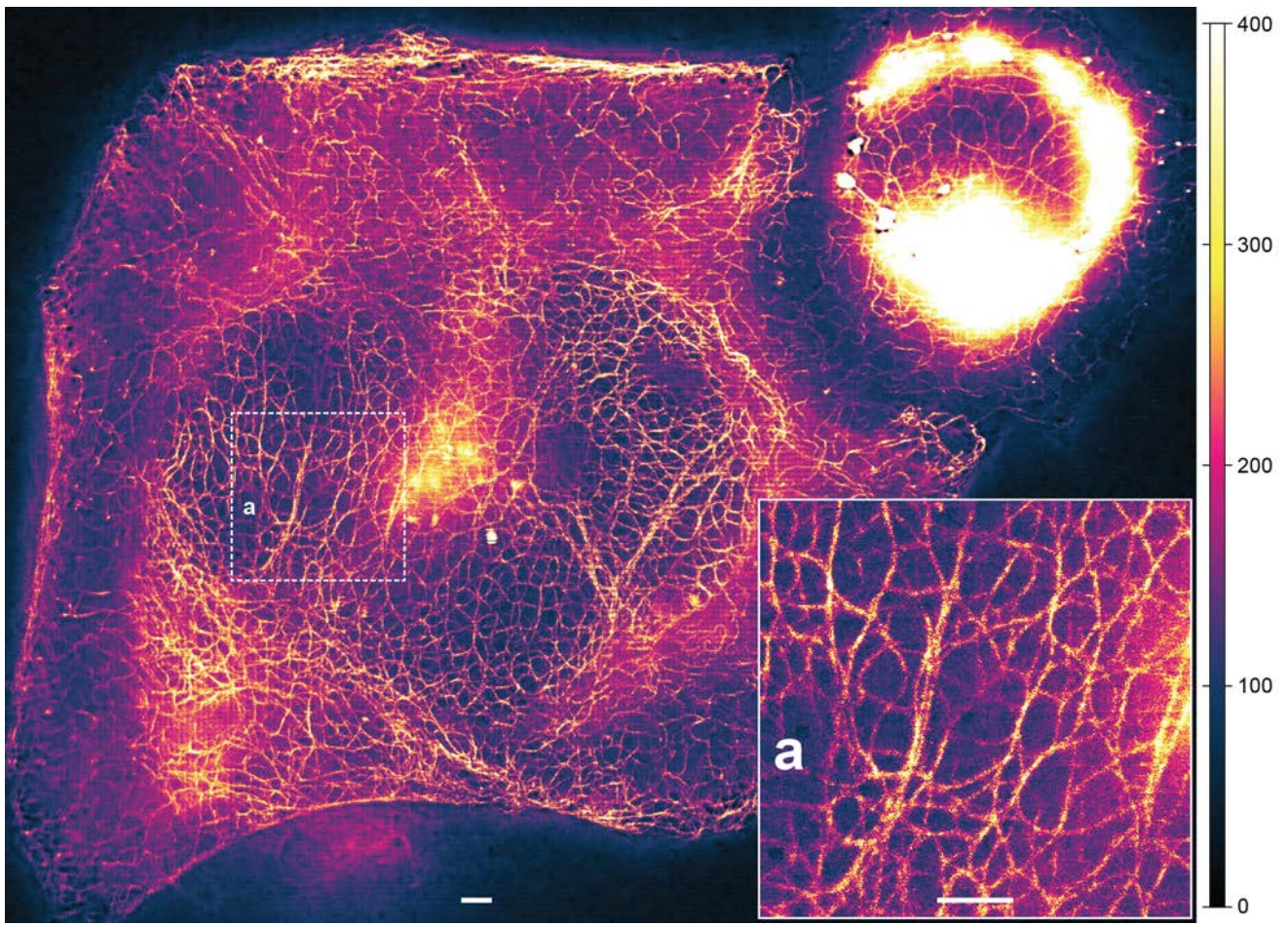

**Supplementary Figure 8 | Parallelized RESOLFT analysis with local pinholes.** The image brightness was saturated at a maximum of 400 and negative values were ignored for clarity (peak brightness 8'090). (**a**) 2.5× magnified region. Scale bars: 2 μm. Displayed fields of view: 78×61 μm$^2$, 11×11 μm$^2$ (**a**).

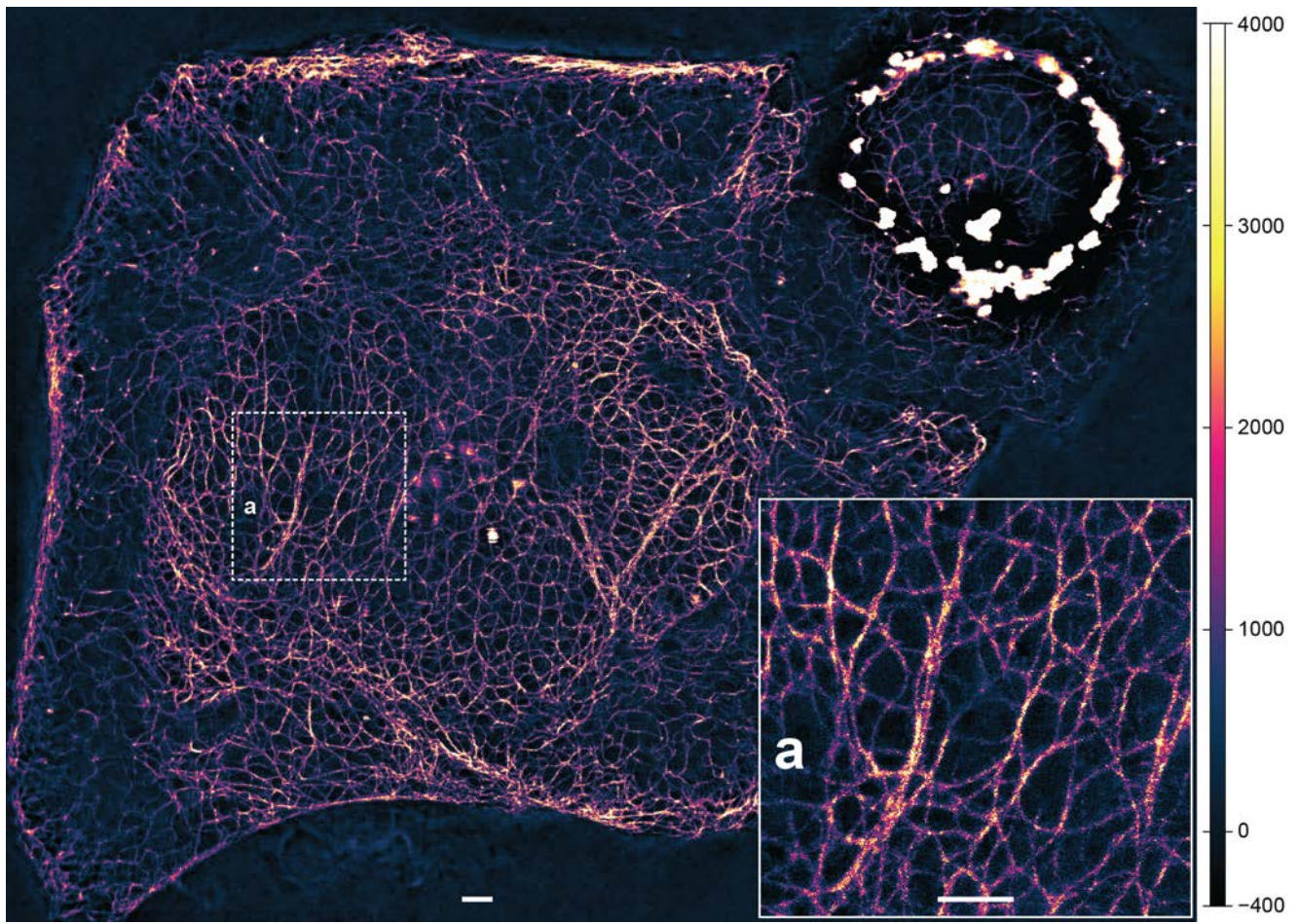

**Supplementary Figure 9 | Background elimination by spatial bandpass filtering of the acquired images.** The RESOLFT image brightness was saturated at a maximum of 4'000 for clarity (peak brightness 130'000). (**a**) 2.5× magnified region. Scale bars: 2 μm. Displayed fields of view: 78×61 μm$^2$, 11×11 μm$^2$ (**a**).

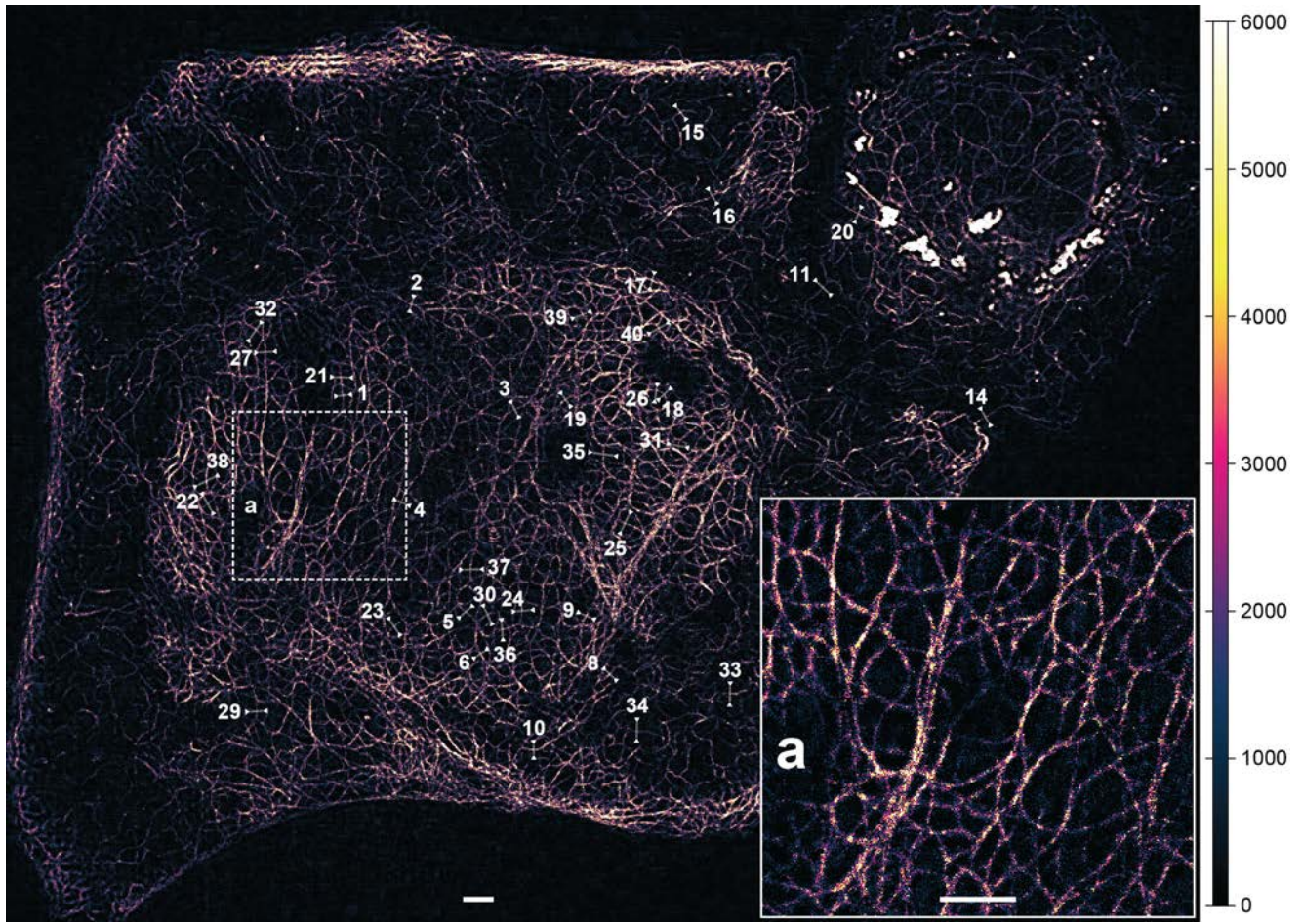

**Supplementary Figure 10 | In-focus image obtained by the model-based RESOLFT image reconstruction.** The image brightness was saturated at a maximum of 6'000 for clarity (peak brightness 164'000). Line profiles are indicated for **Supplementary Figure 11**. (**a**) 2.5× magnified region. Scale bars: 2 μm.

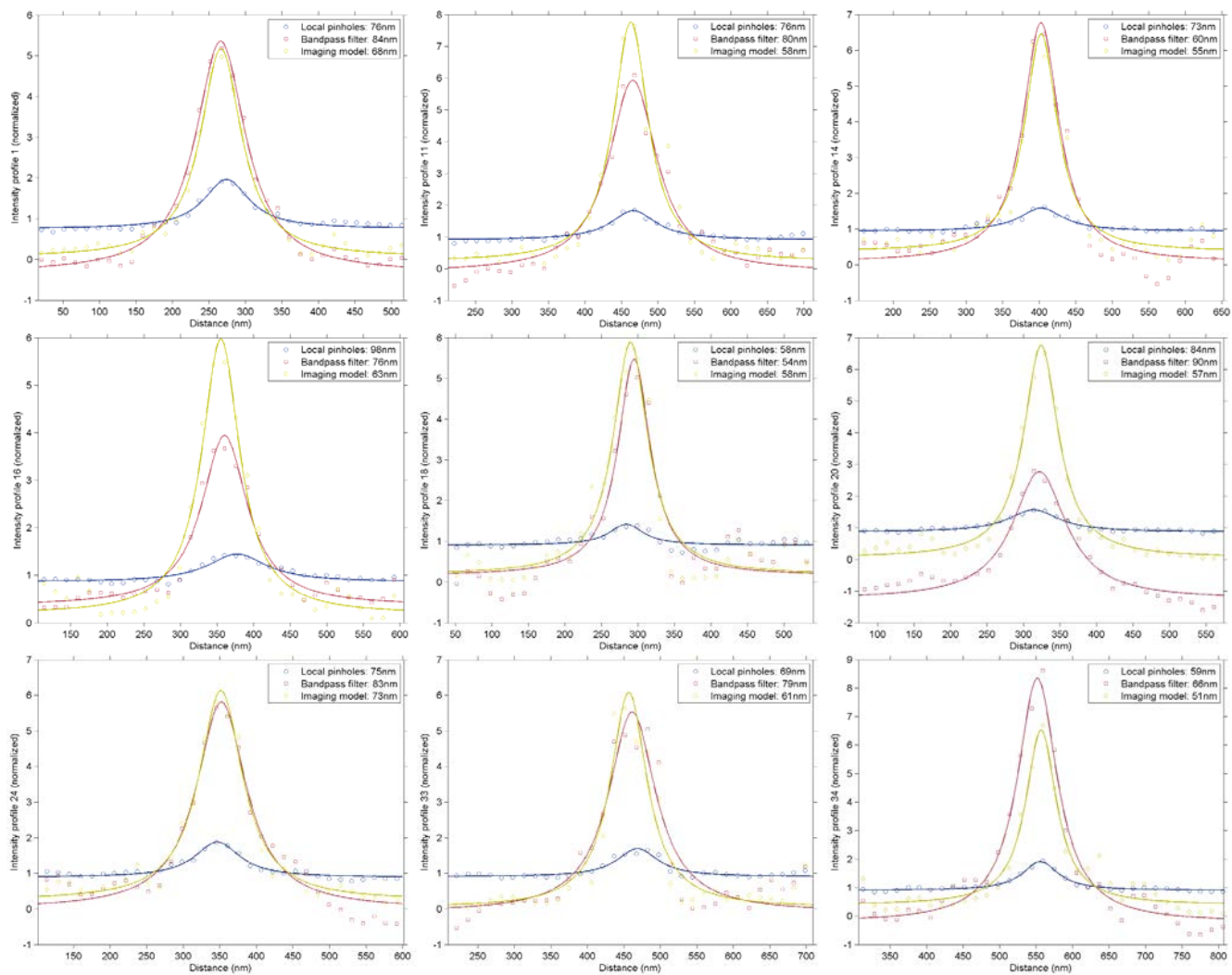

**Supplementary Figure 11 | Line profiles across keratin19 filaments in Supplementary Figure 8–10.** The intensity line profiles (dots) were averaged over a width of 5 pixels (130 nm) for the different RESOLFT analysis methods and then normalized by the averages of their absolute values. The positions of the line profiles are indicated in **Supplementary Figure 10**. Fits to a Lorentz function are shown as solid lines and the corresponding FWHM values are reported. By looking for clearly visible but likely single tubulin filaments, we selected 20 profiles in the local pinholes RESOLFT reconstruction and another 20 profiles in the imaging model RESOLFT reconstruction. From these 40 profiles we excluded four near-duplicates. The model-based analysis performed best with nine profiles (25%) measured at 51–62 nm FWHM and a median FWHM of 76 nm. For the bandpass filtered analysis, nine profiles were measured at 54–80 nm FWHM, the median was 94 nm. The corresponding values for the local pinholes analysis were 58–84 nm FWHM and 104 nm.

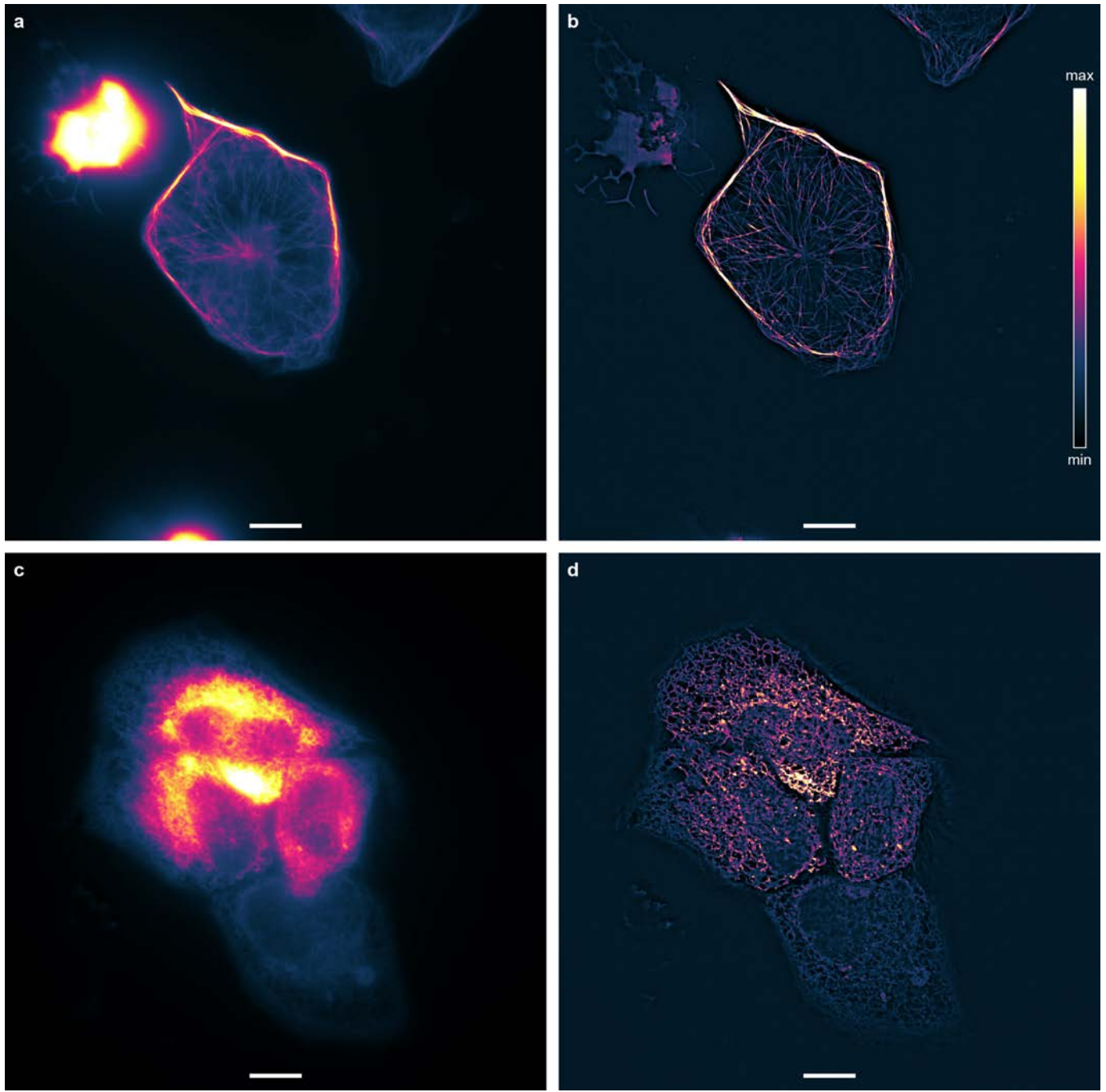

**Supplementary Figure 12 | Full images for magnified regions in Fig. 2k–n.** Wide-field images (**a**,**c**) and RESOLFT reconstruction with bandpass filtering (**b**,**d**) of HeLa cells expressing MAP2–Dreiklang (**a**,**b**), imaged with 12×12 scan steps of 30 nm step length, and KDEL–Dreiklang (**c**,**d**), imaged with 8×8 scan steps of 45 nm. Displayed is the acquired field of view: 104×104 µm$^2$. Scale bars: 10 µm.

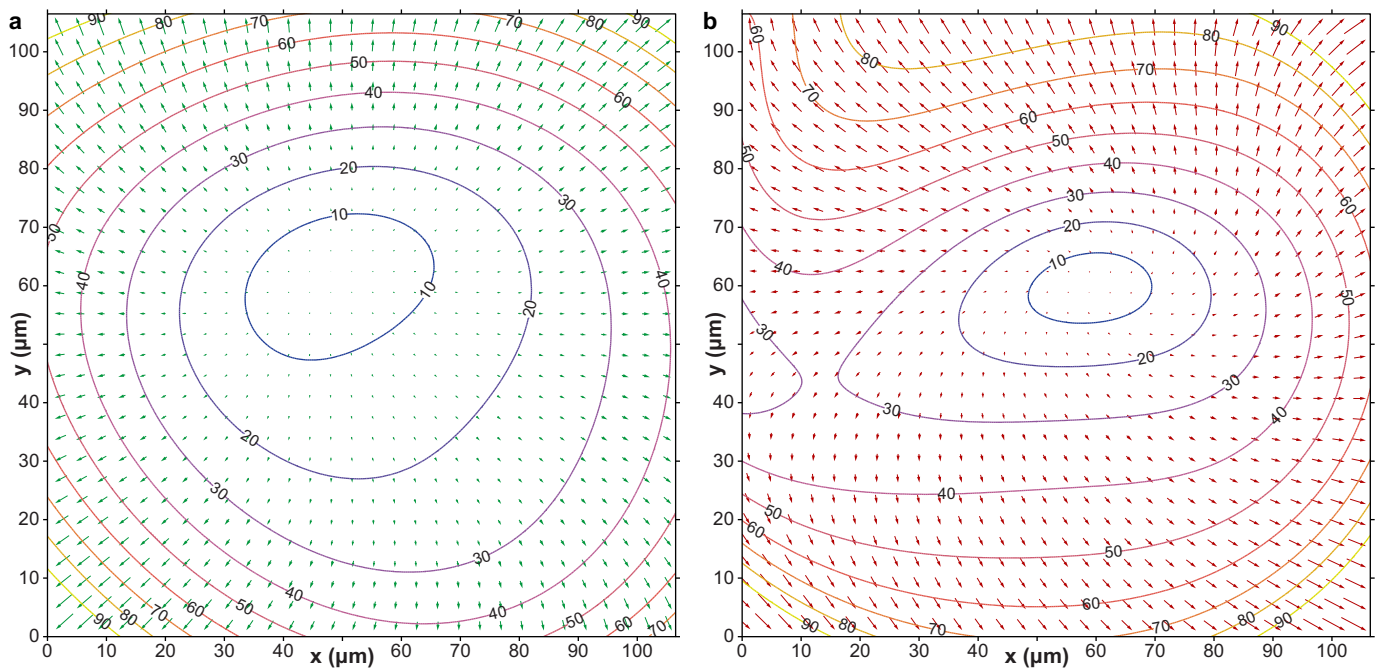

**Supplementary Figure 13 | Mismatches of illumination patterns estimated by images of fluorescent layers.** The contour lines show the relative shift of the fluorescence peak positions of the 488 nm (**a**) and 592 nm (**b**) illuminations versus the 405 nm illumination in percent of the average horizontal and vertical pattern period. The arrows point at every 10$^{th}$ peak of the 405 nm illumination pattern and indicate the distance from the corresponding illumination peak of the other patterns. The lengths of the arrows are magnified 10× to match the increased peak spacing. We deem acceptable horizontal and vertical mismatches of up to 30–40% and, respectively, up to 45–55% along the diagonal directions.

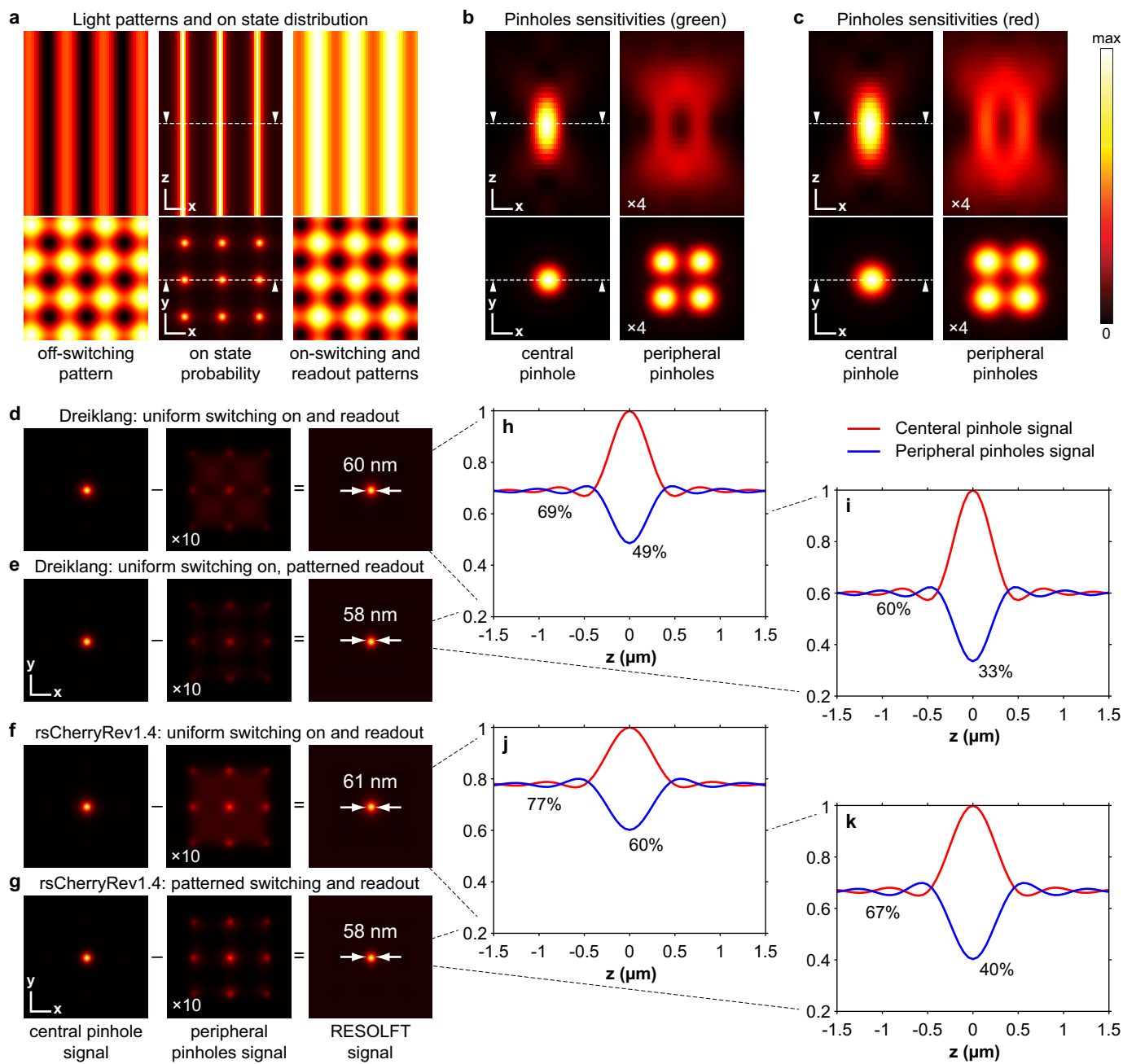

**Supplementary Figure 14 | Calculated lateral resolution and background rejection with and without patterned activation and readout.**
(**a**) Cross-sections of the off-switching illumination pattern ($\Lambda = 360$ nm) (left); potential population of fluorophores in the on state after switching off (center); and cross-sections of the on-switching and readout illumination patterns, shifted half a period with respect to the off-switching pattern (right). (**b,c**) Cross-sections of the fluorescence detection PSF by a small pinhole at a null (left) or four pinholes around the null (right) for a fluorescence wavelength of 520 nm for Dreiklang **b** and 610 nm for rsCherryRev1.4 **c**. (**d**–**g**) Effective imaging PSF (right) as obtained by measuring the central pinhole signal (left) and subtracting the peripheral pinholes signals (center) for fluorophores in the focal plane. Effective PSFs upon imaging Dreiklang and rsCherryRev1.4 with and without patterned on-switching and readout are illustrated and their lateral FWHMs are indicated. (**h**–**k**) Corresponding normalized axial responses of the central and peripheral pinholes for fluorophores located in a thin layer at different axial distances from the focal plane. The background levels and the minimum peripheral signals at the focal plane are indicated. The signals of the peripheral pinholes are scaled up 4× (**b,c**) or 10× (**d**–**g**) for clarity. Scale bars (axes indicators): 200 nm.

# Supplementary Methods

## Optical setup

We adapted the wide-field RESOLFT instrument by Abberior Instruments, whose original optical setup is largely based on setup described in [1]. The modified RESOLFT microscope is shown in **Supplementary Figure 1**. An Olympus IX-83 microscope body was used and customized at its side port. Three continuous-wave laser beams of 405 nm, 488 nm and 592 nm wavelengths were combined by clean-up filters and dichroic mirrors and coupled into a polarization-maintaining single-mode fiber. An achromatic lens $L_1$ collimated the beams at the fiber output to a beam diameter of about 12 mm. An achromatic half-wave retarder plate ($\lambda/2$) and polarizing beam splitters (PBS) sent the beams at equal powers through two binary phase gratings ($G_1$ and $G_2$). The phase gratings diffracted the s-polarized beams primarily into the first diffraction orders. These were focused by the achromatic lens $L_2$ and selected by an order selection mask (OS) to cancel stray light and the order zero. The achromatic lenses $L_3$ and $L_4$ relayed the illumination beams on a tip-tilt piezo scan mirror. The custom-made dichroic mirror $D_1$ transmitted the wavelength ranges of 350–430 nm, 480–490 nm and 588–598 nm. The achromatic lens $L_5$ and the microscope tube lens relayed the beams into the entrance pupil of the microscope objective. Thereby, 50× demagnified images of the phase gratings were produced at the objective's focal plane. Hence, the sample was illuminated with square grids of 360 nm periods. When imaging Dreiklang, we switched the protein's fluorescence on with a homogeneous illumination of 365 nm wavelength provided by a LED source and injected via the dichroic mirror $D_3$. The fluorescence was collected by the microscope objective and sent back via the scan mirror to the main dichroic mirror $D_1$. The lenses $L_5$ and $L_4$ and the achromatic lenses $L_6$ and $L_7/L_8$ relayed the intermediate image onto the sensors of two scientific CMOS cameras. The wavelength ranges of 440–470 nm, 500–575 nm and 610–750 nm were reflected by $D_1$ and further split by the dichroic mirror $D_2$ into two colour channels. The dichroic mirrors and the bandpass filters $F_1$ and $F_2$ selected the fluorescence at 500–575 nm and 610–680 nm wavelengths.

## RESOLFT image reconstruction

The raw wide-field images were analysed in three major steps:

1. Determination of the signal origins corresponding to the positions of the intensity minima ("nulls") of the switching off illumination pattern.
2. Estimation of the in-focus fluorescence emission from the regions of the nulls.
3. Reattribution of the fluorescence emission to confined regions at the nulls.

We improved the image reconstruction over the previously published method [1] to (i) more accurately identify the positions of the nulls by accommodating distortions of the illumination pattern and/or its images, and to (ii) better reject the fluorescence emission from out-of-focus features. The analysis can exclude sensor pixels with high dark current or excessive read noise as identified in dark pictures. The new analysis algorithms are detailed below.

### 1. Signal origins

Whenever possible, the set of positions $X = \{(x_{mn}, y_{mn})\}$ of the nulls were extracted from the raw images of a measurement (**Supplementary Figure 2**). For samples with very sparse features, an extra calibration measurement of a thin layer of fluorescent proteins was performed shortly prior to the image acquisition of the sample.

We assume that the nulls are located on an approximately regular square grid aligned approximately with the camera sensor along its horizontal $x$ and vertical $y$ directions. The nulls are projected at the same sensor positions for the image frames captured during a measurement. Therefore, the raw images of an entire acquisition were summed to obtain $I(x,y)$, an image of the average fluorescence emission in all unit cells across the field of view (**Supplementary Figure 2a**).

The grid periods $\Lambda_x$ and $\Lambda_y$ were estimated by identifying the spatial frequency of the major peak in the horizontal and vertical spatial power spectra of the sum image (**Supplementary Figure 2b**). These spectra were calculated with the Fourier transform with a frequency sampling fine enough to achieve a precision of about three digits in the period estimates (we used 128× oversampling). To account for potential misalignments and distortions of the illumination grid, the power spectra were calculated on moving averages spanning 30 pixels along the opposite direction (**Supplementary Figure 2a**). The periods were then estimated by a weighted average on the peak spatial frequencies, where the weights were proportional to the peak spectral power density (**Supplementary Figure 2b**).

Next, local offsets of the signal origins were determined by correlating the sum image $I(x,y)$ with the estimated $x$ and $y$ grid in neighbourhoods of 20×20 periods extent (**Supplementary Figure 2c**). For the horizontal position offsets, the sum image was multiplied by $F_x(x) = \exp(2i\pi x/\Lambda_x)$ and correlated with a pyramidal mask $W(x,y) = \max(0; 1 - |x/10\Lambda_x|) \times \max(0; 1 - |y/10\Lambda_y|)$:

$$C_x(x,y) = (I(x,y)F_x(x)) * W(x,y) \tag{1}$$

Pixel by pixel, the phase $\phi_x(x,y) = \arg C_x(x,y)$ of this moving correlation (**Supplementary Figure 2d**) indicates the position offset $\Delta x(x,y) = \phi_x(x,y)\Lambda_x/2\pi$ of the local null with respect to the grid $F_x(x)$. In order to locate the signal origins unambiguously and reduce the phase noise, the phase $\phi_x(x,y)$ was unwrapped (c.f. algorithm by Herráez et al. [2]) and the resulting phase map $\Phi_x(x,y)$ was low-pass filtered by the first 25 fringe Zernike polynomials with weights $|C_x(x,y)|$ (**Supplementary Figure 2e**). The Zernike polynomials allowed to extrapolate the phase map in the border region of partial overlap of $I(x,y)F_x(x)$ and $W(x,y)$. The vertical position offsets were found analogously using the grid pattern $F_y(y) = \exp(2i\pi y/\Lambda_y)$ along the vertical direction to estimate $\Delta y(x,y)$.

The signal origins $X$ (**Supplementary Figure 2f**) were then obtained by shifting the initial positions $(x_{mn,0}, y_{mn,0}) = (m\Lambda_x, n\Lambda_y)$, with integer $m$ and $n$, by the position offsets $(\Delta x, \Delta y)(x_{mn,0}, y_{mn,0})$. To reduce the error in case of significant offsets, the position estimates were updated several times with only a fraction of the total offset:

$$(x_{nm,j}, y_{mn,j}) = (x_{mn,j-1}, y_{mn,j-1}) + \frac{(\Delta x, \Delta y)(x_{mn,j-1}, y_{mn,j-1})}{J} \quad \forall j \in \{1,2,...,J\} \tag{2}$$

Within regions of sample features, this local refinement of the position estimates allowed to accommodate for errors of up to a few percent in the estimation of the grid periods. Moreover, it accounted for small misalignments of the illumination patterns with the camera's sensor axes. Simulations showed that the signal origins were reliably found with less than one tenth of a period position error for sum images with SNR > 2.

## 2. Signal retrieval

### Signal and background estimation by local pinholes

Signals were locally integrated using Gaussian-weighted masks to include neighbouring pixels at the nulls. For each scan step, the local background at each null was determined by 80% of the mean of the signals obtained at the four nearest-neighbour positions diagonally in between the nulls, and then subtracted. For further details of the computational image reconstruction see [1].

### Background elimination by spatial bandpass filtering

**Supplementary Figure 3** illustrates the image analysis. Each raw image (**Supplementary Figure 3a**) was smoothed with a Gaussian filter of $\text{FWHM} = (\Lambda_x + \Lambda_y)/(4\sqrt{2})$ for reducing noise and the Laplacian was taken from the smoothed images to eliminate the background (**Supplementary Figure 3b**). These bandpass-filtered raw images were then integrated in small regions (pinholes) at the nulls' positions $X$ to estimate the fluorescence emitted there (**Supplementary Figure 3d**). The pinholes were determined by the positive signals in 3×3 camera pixels neighbourhoods centred at the nulls (**Supplementary Figure 3c**). In order to equalize their sensitivities, the pinholes' signals were determined by processing a simulated image of equally bright emitters at the nulls.

This Laplacian-of-Gaussian filtering eliminated background efficiently but introduced some artefacts due to partially overlapping images of adjacent nulls (observe for instance negative values in **Supplementary Figure 3b,d**). Therefore, images of fluorescence emitters in the focal section at single nulls were simulated (described in the next section) and processed to estimate the point spread functions including the image analysis (**Supplementary Figure 3e**). The artefacts were then reduced by redistributing the pinhole signals (**Supplementary Figure 3c**) such that the squared difference to the observed bandpass filtered image (**Supplementary Figure 3b**) was minimized (**Supplementary Figure 3f**).

### Model-based maximum likelihood estimation of signals

Given the parameters of the microscope objective and the illumination optics, the illumination patterns in a lateral plane of the sample were estimated with the fast focus field calculation tool [3] (**Supplementary Figure 4**).

The on- and off-switching and the emission of the fluorescent protein is illustrated in **Supplementary Figure 4a** and was estimated as described in the RESOLFT section in reference [4]. We assumed to switch on up to 80% of the proteins ($n_{on} < 1.6$).[1] We assumed an off-switching illumination dose to reach a five-fold improvement in lateral resolution ($n_{off} < 25$), which corresponded to 60–70 nm FWHM of the regions with emissive fluorescent proteins. We further assumed to switch off up to 80% during read-out of the fluorescence ($n_{det} < 1.6$) for rsCherryRev1.4 and up to 25% ($n_{det} < 0.3$) for Dreiklang.[2]

The detection point spread function (PSF) was estimated with the calculation described by Leutenegger and Lasser [5]. The detection PSF was first calculated in lateral planes with a resolution of 2 nm over an axial range of 2 μm in steps of 10 nm. Neighbouring lateral planes were then grouped into sections such that the areas of the PSFs in adjacent sections quadrupled (**Supplementary Figure 4b**). The planes of the focal section were grouped such that the focal PSF area increased by 42% with respect to the smallest PSF area in the focal plane. Thus, a set of significantly different PSF sections $PSF_i(x, y)$ was obtained, centered at axial positions $z_i$ and spanning an axial range $\Delta z_i$. In particular, we used $z \in \{110, 305, 680, 1390\}$ nm and $\Delta z \in \{250, 120, 650, 780\}$ nm for modelling the focal section at the glass cover slip and three neighbouring sections in the sample. The fluorescence signal from a unit cell – calculated as outlined in the previous paragraph – was then convoluted with the section PSFs to estimate the images of features in these sections.

The image formation model $\vec{y} = \boldsymbol{M}\vec{c} + \vec{\varepsilon}$ consisted of a sparse matrix $\boldsymbol{M} = (M_{mn})$, whose element $M_{mn}$ was the contribution to pixel $m$ by the PSF of the $n^{th}$ coefficient. The elements $(c_n)$ of the coefficient column vector $\vec{c}$ were the estimated fluorescence signals at the positions $\boldsymbol{X}$ in the sections $i$. The measured image in column vector $\vec{y}$ was thus obtained by the matrix product $\boldsymbol{M}\vec{c}$ plus shot noise and read-out noise $\vec{\varepsilon}$. In order to build the model $\boldsymbol{M}$ column by column, the section PSFs were placed at the lateral positions $\boldsymbol{X}$ and binned into the camera pixels. Thereby, sub-pixel shifts of the nulls' positions with respect to the camera pixels were taken into account.

A conjugate gradient solver was used to find the coefficients $\vec{c}$ maximizing the Poisson likelihood (photo-electron statistics) that the estimated image $\boldsymbol{M}\vec{c}$ corresponds to the measured raw image $\vec{y}$. This fitting procedure eliminated the blurred signals of defocused features. However, only the coefficients of the focal section could be estimated with high spatial resolution as the others showed significant artefacts due to the overlap of the signals of adjacent nulls. The iterative fitting took about an hour per 100 raw images of a RESOLFT image acquisition.

### 3. Signal reattribution

The retrieved signals were combined into the RESOLFT image by placing each signal at its new pixel position in the final image. Photo-bleaching was accounted for by weighing the signals with the inverse of the average signals in each raw image, whereby a second-order polynomial decay of the average signals was assumed to avoid artefacts due to the sample structure.

## Wide-field image reconstruction

In order to compare the RESOLFT images with wide-field images, we combined the raw images as follows. We up-sampled the raw images to match the spatial sampling of the RESOLFT image and shifted each raw image to its position within the unit cell of the illumination patterns. These steps were performed by zero-padding and phase-shifting the spatial spectra of the raw images. The resampled raw images were then summed with weights inversely proportional to the average signals in each image to account for photo-bleaching (see 3. Signal reattribution). This wide-field image was further processed by a Lucy-Richardson deconvolution, where we used the theoretical diffraction-limited detection efficiency as PSF.

# Supplementary References

1. Chmyrov A, *et al*. Nanoscopy with more than 100,000 'doughnuts'. *Nature Methods* **10**, 737–740 (2013).

---

[1] Notation as in review [4]: $n_{on}$ designates the number of transitions to the bright on state a fluorescent protein could perform if it would remain in its dark off state during the on-switching illumination pulse. After switching on, the off state is depopulated to a fraction of $\exp(-n_{on})$.
[2] Dreiklang switches off slowly upon illumination at 488 nm wavelength.